\documentclass{llncs}

\usepackage[german,english]{babel}
\usepackage{times,theorem}
\usepackage{latexsym,color,graphics}
\usepackage{diagrams}
\usepackage{url}
\usepackage{epsfig}
\usepackage{amssymb}
\usepackage{amsmath}
\usepackage{amsfonts}

\begingroup
\catcode`\~=11
\gdef\urltilde{\lower 0.6ex\hbox{~}}
\endgroup

\newcommand{\A}{\mathcal{A}} 
 \newcommand{\D}{\mathcal{D}}
\newcommand{\E}{\mathcal{E}} 
 
\newcommand{\I}{\mathcal{I}} 
\newcommand{\K}{\mathcal{K}} \renewcommand{\L}{\mathcal{L}}
\newcommand{\M}{\mathcal{M}} \newcommand{\N}{\mathcal{N}}
 \renewcommand{\P}{\mathcal{P}}
 \newcommand{\R}{\mathcal{R}}
 \newcommand{\T}{\mathcal{T}}
 \newcommand{\V}{\mathcal{V}}
\newcommand{\W}{\mathcal{W}}

\title{First-order Logic: Modality and Intensionality}
\author{Zoran Majki\'c}
\authorrunning{Zoran Majki\'c}
\institute{International Society for Research in Science and Technology \\
PO Box 2464 Tallahassee, FL 32316 - 2464 USA\\
\email{majk.1234@yahoo.com}\\ http://zoranmajkic.webs.com/}
\authorrunning{Zoran Majki\'c}

\newtheorem{theo}{Theorem}
\newtheorem{propo}{Proposition}
\newtheorem{coro}{Corollary}

\newcount\pdfoutput
\begin{document}


\maketitle

\begin{abstract}
Contemporary use of the term 'intension' derives from the
traditional logical Frege-Russell's doctrine that an idea (logic
formula) has both an extension and an intension. Although there is
divergence in formulation, it is accepted that the extension of an
idea  consists of the subjects to which the idea applies, and the
intension consists of the attributes implied by the idea. From the
Montague's point of view, the meaning of an idea  can be considered
as particular extensions in different possible worlds.
\\In this paper we analyze the minimal intensional semantic
enrichment of the syntax of the FOL language, by unification of
different views: Tarskian extensional semantics of the FOL, modal
interpretation of quantifiers, and a derivation of the Tarskian
theory of truth from unified semantic theory based on a single
meaning relation. We show that not all modal predicate logics are
intensional, and that an equivalent modal Kripke's interpretation of
logic quantifiers in FOL results in a particular pure extensional
modal predicate logic (as is the standard Tarskian semantics of the
FOL). This minimal intensional enrichment is obtained by adopting
the theory of properties, relations and propositions (PRP) as the
universe or domain of the FOL, composed by particulars and
universals (or concepts), with the two-step interpretation of the
FOL that eliminates the  weak points of the Montague's intensional
semantics.  Differently from the Bealer's intensional FOL, we show
that it is not necessary the introduction of the intensional
abstraction in order to obtain the full intensional properties of
the FOL.\\ Final result of this paper is represented by the
commutative homomorphic  diagram that holds in each given possible
world of this new intensional FOL, from the free algebra of the FOL
syntax, toward its intensional algebra of concepts, and,
successively, to the new extensional relational algebra (different
from Cylindric algebras), and we show that it corresponds to the
Tarski's interpretation of the standard extensional FOL in this
possible world.
\end{abstract}


\section{Introduction}
The simplest aspect of an expression's meaning is its extension. We
can stipulate that the extension of a sentence is its truth-value,
and that the extension of a singular term is its referent. The
extension of other expressions can be seen as associated entities
that contribute to the truth-value of a sentence in a manner broadly
analogous to the way in which the referent of a singular term
contributes to the truth-value of a sentence. In many cases, the
extension of an expression will be what we intuitively think of as
its referent, although this need not hold in all cases, as the case
of sentences illustrates. While Frege himself is often interpreted
as holding that a sentence's referent is its truth-value, this claim
is counterintuitive and widely disputed. We can avoid that issue in
the present framework by using the technical term 'extension'. In
this context, the claim that the extension of a sentence is its
truth-value is a stipulation.\\
'Extensional' is most definitely a technical term. Say that the
extension of a name is its denotation, the extension of a predicate
is the set of things it applies to, and the extension of a sentence
is its truth value. A logic is extensional if coextensional
expressions can be substituted one for another in any sentence of
the logic "salva veritate", that is, without a change in truth
value. The intuitive idea behind this principle is that, in an
extensional logic, the only logically significant notion of meaning
that attaches to an expression is its extension. An intensional
logics is exactly one in which substitutivity salva veritate fails
for some of the sentences of the logic.\\
 In "$\ddot{U}ber~ Sinn~
und~ Bedeutung$", Frege concentrated mostly on the senses of names,
holding that all names have a sense. It is natural to hold that the
same considerations apply to any expression that has an extension.
Two general terms can have the same extension and different
cognitive significance; two predicates can have the same extension
and different cognitive significance; two sentences can have the
same extension and different cognitive significance. So general
terms, predicates, and sentences all have senses as well as
extensions. The same goes for any expression that has an extension,
or is a candidate for extension.\\ The distinction between
intensions and extensions is important, considering that extensions
can be notoriously difficult to handle in an efficient manner. The
extensional equality theory of predicates and functions under
higher-order semantics (for example, for two predicates with the
same set of attributes $p = q$ is true iff these symbols are
interpreted by the same relation), that is, the strong equational
theory of intensions, is not decidable, in general. For example, in
the second-order predicate calculus and Church's simple theory of
types, both under the standard semantics, is not even
semi-decidable. Thus, separating intensions from extensions makes it
possible to have an equational theory over predicate and function
names (intensions) that is separate from the extensional equality of
relations and functions. \\
The first conception of intensional entities (or concepts) is built
into the \emph{possible-worlds} treatment of Properties, Relations
and Propositions (PRP)s. This conception is commonly attributed to
Leibniz, and underlies Alonzo Church's alternative formulation of
Frege's theory of senses ("A formulation of the Logic of Sense and
Denotation" in Henle, Kallen, and Langer, 3-24, and "Outline of a
Revised Formulation of the Logic of Sense and Denotation" in two
parts, Nous,VII (1973), 24-33, and VIII,(1974),135-156). This
conception of PRPs is ideally suited for treating the
\emph{modalities} (necessity, possibility, etc..) and to Montague's
definition of intension of a given virtual predicate
$\phi(x_1,...,x_k)$ (a FOL open-sentence with the tuple of free
variables $(x_1,...x_k)$) as a mapping from possible worlds into
extensions of this virtual predicate. Among the possible worlds we
distinguish the \emph{actual} possible world. For example if we
consider a set of predicates of a given Database
and their extensions in different time-instances, the actual possible world is identified by the current instance of the time.\\
The second conception of intensional entities is to be found in in
Russell's doctrine of logical atomism. On this doctrine it is
required that all complete definitions of intensional entities be
finite as well as unique and non-circular: it offers an
\emph{algebraic} way for definition of complex intensional entities
from simple (atomic) entities (i.e., algebra of concepts),
conception also evident in Leibniz's remarks. In a predicate logics,
predicates and open-sentences (with free variables) expresses
classes (properties and relations), and sentences express
propositions. Note that classes (intensional entities) are
\emph{reified}, i.e., they belong to the same domain as individual
objects (particulars). This endows the intensional logics with a
great deal of uniformity, making it possible to manipulate classes
and individual objects in the same language. In particular, when
viewed as an individual
object, a class can be a member of another class.\\
The standard semantics of First-order Logic (FOL) are Tarski style
models, which are extensional. In this respect, FOL is extensional.
But the open question is if it is possible to obtain also an
intensional semantics of FOL such that the Tarski's extensions of
its expressions are equal to extensions of concepts (intensional
entities) of the same FOL expressions in the actual possible world. \\
In what follows we  denote by $B^A$ the set of all functions from
$A$ to $B$, and by $A^n$ a n-folded cartesian product $A \times
...\times A$ for $n \geq 1$. By $f, t$ we denote  empty set
$\emptyset$ and singleton set $\{<>\}$ respectively (with the empty
tuple $<>$ i.e. the unique tuple of 0-ary relation), which may be
thought of
 as falsity $f$ and truth $t$, as those used  in the relational algebra.
 For a given domain $\D$ we define that $\D^0$ is a singleton set $\{<>\}$, so that $\{f, t\} = \P(\D^0)$, where $\P$ is the powerset operator.
%
%

\textbf{First-order Logic (FOL):}
 We will shortly introduce the syntax of
the FOL language $\L$, and its extensional semantics based on
Tarski's interpretations, as follows:
 \begin{definition} \label{def:FOL}
 The syntax of the First-order Logic language $\L$  is as follows:\\
 Logic operators $(\wedge,  \neg, \exists)$ over bounded lattice  of
   truth values  $\textbf{2} = \{f,t\}$, $f$ for falsity and $t$ for truth; Predicate letters $p_1^{k_1},p_2^{k_2},...$ with a given arity $k_i\geq 1$,
   $i = 1,2,..$    in
  $P$; Functional letters $f_1^{k_1},f_2^{k_2},...$ with a given arity $k_i\geq 1$ in $F$
 (language constants $c,d,...$ are considered as particular case of nullary
 functional letters); Variables $x,y,z,..$ in $\V$, and punctuation
 symbols (comma, parenthesis).\\
 With the following simultaneous inductive definition of \emph{term} and
 \emph{formula}:\\
   1. All variables and constants  are terms.\\
   2. If $~t_1,...,t_k$ are terms and $f_i^k \in F$ is a k-ary functional symbol then $f_i^k(t_1,...,t_k)$ is a
   term, while $p_i^k(t_1,...,t_k)$ is a formula
 for a k-ary predicate letter $p_i^k \in P$.\\
   3. If $\phi$ and $\psi$ are formulae, then $(\phi \wedge \psi)$, $\neg \phi$,  and $(\exists x_i) \phi$ for $x_i \in \V$ are
   formulae.\\
   An interpretation (Tarski) $I_T$ consists in a non empty domain
   $\D$ and a mapping that assigns to any predicate letter $p_i^k \in
   P$ a relation $R = I_T(p_i^k) \subseteq \D^k$, to any functional letter $f_i^k \in
   F$ a function $I_T(f_i^k): \D^k \rightarrow \D$, or, equivalently, its graph relation $R = I_T(f_i^k)\subseteq \D^{k+1}$ where the $k+1$-th column is
   the resulting function's value, and to each
   individual constant $c \in F$ one given element $I_T(c) \in
   \D$.\\
   A Predicate Logic $\L_P$ is a subset of the FOL without the
   quantifier $\exists$.
\end{definition}
\textbf{Remark:} The \emph{propositional} logic can be considered as
a particular case of Predicate logic when all symbols in $P$ are
nullary, that is a set of propositional symbols, while $F$, $\D$,
and $\V$ are empty sets. By considering that $\D^0 = \{<>\}$ is a
singleton set, then for any $p_i \in P$, $I_T(p_i) \subseteq \D^0$,
that is, $I_T(p_i) = f$ (empty set) or $I_T(p_i) = t$ (singleton set
$\{<>\})$. That is, $I_T$ becomes an  interpretation $I_T:P
\rightarrow \textbf{2}$ of this  logic, which can be homomorphically
extended to all  formulae in the unique standard way.
 \\$\square$\\
 In a formula $(\exists x)\phi$, the formula
$\phi$ is called "action field" for the quantifier $(\exists x)$. A
variable $y$ in a formula $\psi$ is called bounded variable iff it
is the variable of a quantifier $(\exists y)$ in $\psi$, or it is in
the action field of a quantifier $(\exists y)$ in the formula
$\psi$. A variable $x$ is
free in $\psi$ if it is not bounded. \\
The universal quantifier is defined by $\forall = \neg \exists
\neg$. Disjunction and implication are expressed by
 $\phi \vee \psi = \neg(\neg \phi \wedge \neg \psi)$, and $\phi \Rightarrow \psi = \neg \phi \vee \psi$.
 In FOL with the
identity $\doteq$, the formula $(\exists_1 x)\phi(x)$ denotes the
formula $(\exists x)\phi(x) \wedge (\forall x)(\forall y)(\phi(x)
\wedge \phi(y)  \Rightarrow (x \doteq y))$. \\
  We can introduce the sorts in
order to be able to assign each variable $x_i$ to a sort $S_i
\subseteq \D$ where $\D$ is a given domain for the FOL
 (for example, for natural
numbers, for reals,  for dates, etc.. as used for some attributes in
database relations).
 An assignment $g:\V \rightarrow \D$ for variables in $\V$ is
applied only to free variables in terms and formulae. If we use
sorts for variables, then for each sorted variable $x_i \in \V$ an
assignment $g$ must satisfy the auxiliary condition $g(x_i) \in
S_i$. Such an assignment $g \in \D^{\V}$ can be recursively uniquely
extended into the assignment $g^*:\T \rightarrow \D$, where $\T$
denotes the
set of all terms, by:\\
1. $g^*(t) = g(x) \in \D$ if the term $t$ is a variable $x \in
\V$.\\
2. $g^*(t) = I_T(c) \in \D$ if the term $t$ is a constant $c \in
F$.\\
3. if a term $t$ is $f_i^k(t_1,...,t_k)$, where $f_i^k \in F$ is a
k-ary functional symbol and $t_1,...,t_k$ are terms, then
$g^*(f_i^k(t_1,...,t_k)) = I_T(f_i^k)(g^*(t_1),...,g^*(t_k))$ or,
equivalently, in the graph-interpretation of the function,
 $g^*(f_i^k(t_1,...,t_k)) = u$ such that $(g^*(t_1),...,g^*(t_k),u)\\ \in I_T(f_i^k)\subseteq \D^{k+1}$.\\
 In what follows we will use the graph-interpretation for functions
 in FOL like its interpretation in intensional logics.
 We denote by $~t/g~$ (or $\phi/g$) the ground term (or
formula) without free variables, obtained by assignment $g$ from a
term $t$ (or a formula $\phi$), and by  $\phi[x/t]$ the formula
obtained by  uniformly replacing $x$ by a term $t$ in $\phi$.
A \emph{sentence} is a formula having no free variables. \\
 A Herbrand base of a  logic $\L$ is defined by $H = \{
p_i^k(t_1,..,t_k)~|~p_i^k \in P$ and $t_1,...,t_k$ are ground terms
$ \}$. We define  the satisfaction for the logic formulae
in $\L$ and a given assignment $g:\V \rightarrow \D$ inductively, as follows:\\
 If a formula $\phi$ is an atomic formula $p_i^k(t_1,...,t_k)$,
then this assignment $g$ satisfies $\phi$ iff
$(g^*(t_1),...,g^*(t_k)) \in I_T(p_i^k)$;
 $~g$ satisfies $\neg \phi~$ iff it does not satisfy $\phi$;
 $~g$ satisfies $\phi \wedge \psi~$ iff $g$ satisfies $\phi$ and
$g$ satisfies $\psi$;  $~g$ satisfies $(\exists x_i)\phi~$ iff
exists an assignment $g' \in \D^{\V}$ that may differ from $g$ only
for the variable $x_i \in
\V$, and $g'$ satisfies $\phi$.\\
A formula $\phi$ is \verb"true" for a given interpretation $I_T~$
iff $~\phi$ is satisfied by every assignment $g \in \D^{\V}$. A
formula $\phi$ is \verb"valid" (i.e., tautology) iff $~\phi$ is true
for every Tarksi's interpretation $I_T \in \mathfrak{I}_T$.  An
interpretation $I_T$ is a \verb"model" of a set of formulae
$\Gamma~$ iff every formula $\phi \in \Gamma$ is true in this
interpretation. We denote by FOL$(\Gamma)$ the FOL with a set of
assumptions $\Gamma$, and by $\mathfrak{I}_T(\Gamma)$ the subset of
Tarski's interpretations that are models of $\Gamma$, with
$\mathfrak{I}_T(\emptyset) = \mathfrak{I}_T$. A formula $\phi$ is
said to be a \emph{logical consequence} of $\Gamma$, denoted by
$\Gamma \Vdash \phi$, iff $\phi$ is true in all interpretations in
$\mathfrak{I}_T(\Gamma)$. Thus, $ ~\Vdash \phi$ iff $\phi$ is a tautology.\\
 The basic set of axioms of the FOL are that of the propositional logic with two
 additional axioms: (A1) $(\forall x)(\phi \Rightarrow \psi)
 \Rightarrow (\phi \Rightarrow (\forall x)\psi)$, ($x$ does not
 occur in $\phi$ and it is not bound in $\psi$), and (A2) $(\forall
 x)\phi \Rightarrow \phi[x/t]$, (neither $x$ nor any variable in $t$ occurs bound in $\phi$).
For the FOL with identity, we need the \emph{proper} axiom (A3) $x_1
\doteq x_2 \Rightarrow (x_1 \doteq x_3 \Rightarrow x_2 \doteq x_3)$.
We denote by $R_{=}$ the Tarski's interpretation of $\doteq$.
\\The
 inference rules are Modus Ponens and generalization (G) "if $\phi$ is a
 theorem
 and $x$ is not bound in $\phi$, then $(\forall x)\phi$ is a
 theorem".\\
 In what follows any open-sentence, a formula
$\phi$ with non empty tuple of free variables $(x_1,...,x_m)$, will
be called a m-ary
  \emph{virtual predicate}, denoted also by
$\phi(x_1,...,x_m)$. This definition contains the precise method of
establishing the \emph{ordering} of variables in this tuple:
  such an method that will be adopted here is the ordering of appearance, from left to right, of free variables in $\phi$.
   This method of composing the tuple of free variables
  is the unique and canonical way of definition of the virtual predicate from a given formula. The FOL is considered as an extensional logic
because two open-sentences with the same tuple of variables
$\phi(x_1,...,x_m)$ and $\psi(x_1,...,x_m)$ are \verb"equal" iff
they have the \emph{same extension} in a given interpretation $I_T$,
that is iff $I_T^*(\phi(x_1,...,x_m)) = I_T^*(\psi(x_1,...,x_m))$,
where $I_T^*$ is the unique extension of $I_T$ to all formulae, as
follows:\\
1. For a (closed) sentence $\phi/g$ we have that $I_T^*(\phi/g) = t$
 iff $g$ satisfies $\phi$, as recursively defined above.\\
2. For an open-sentence $\phi$ with the tuple of free variables
$(x_1,...,x_m)$ we have that $I_T^*(\phi(x_1,...,x_m)) =_{def}
\{(g(x_1),...,g(x_m))~|~ g \in \D^{\V}$ and $I_T^*(\phi/g) = t
\}$.\\
 It is easy to verify that for a formula $\phi$ with the tuple of free variables $(x_1,...,x_m)$,\\ $~I_T^*(\phi(x_1,...,x_m)/g) = t~~$ iff
 $~~(g(x_1),...,g(x_m)) \in I_T^*(\phi(x_1,...,x_m))$.\\
 This extensional \emph{equality} of virtual predicates can be generalized to the extensional \emph{equivalence}
 when both predicates $\phi, \psi$ has the same set of free variables but their
 ordering in the \emph{tuples} of free variables are not identical:
 such two virtual predicates are equivalent if the extension of the
 first is equal to the proper permutation of columns of the extension of the
 second virtual predicate. It is easy to verify that such an
 extensional equivalence corresponds to the logical equivalence denoted by $\phi \equiv \psi$.\\
 Let $\mathfrak{R} = \bigcup_{k \in \mathbb{N}} \P(\D^k) = \sum_{k\in \mathbb{N}}\P(D^k)$ be the set of all k-ary relations
 over a domain $\D$, where $k \in \mathbb{N} =
\{0,1,2,...\}$. Then, this extensional equivalence between two
relations $R_1, R_2 \in \mathfrak{R}$ with the same arity will be
denoted by $R_1 \approx R_2$, while the extensional identity will be
denoted in the standard way by $R_1 = R_2$.

 \textbf{Predicate/Propositional Multi-modal Logics:}\\
A predicate/propositional multi-modal logic  is a standard
Predicate/Propositional Logic (see Definition \ref{def:FOL})
extended by a number of existential \emph{modal} operators
$\lozenge_i, i \geq 1$. In the standard Kripke semantics each modal
operator $\lozenge_i$ is defined by an accessibility binary relation
$\R_i \subseteq \W \times \W$, for a given set of \emph{possible
worlds} $\W$. A more exhaustive and formal introduction to modal
logics and their Kripke's interpretations can easily be found in the
literature, for example in \cite{BBWo06}. Here  only a short version
will be given, in order to clarify the definitions used in the next
paragraphs.\\
 We define $\N = \{0,1,2,..., n\} \subset \mathbb{N}$
where $n$ is a maximal arity of symbols in the finite set $P \bigcup
F$ of predicate and functional symbols respectively. In the case of
the propositional logics we have that $n = 0$, so that $P$ is a set
of propositional symbols (that are the nullary predicate symbols)
and $F = \emptyset$ is the empty set. Here we will present two
definitions for modal logics, one for the propositional and other
for predicate logics, as is used in current literature:
\begin{definition} \label{def:KripSemProp}\textsc{Propositional multi-modal logic}:\\
 We denote by $\M = (\W, \{$$
{\R}_i \},  I_K)$ a multi-modal  Kripke's interpretation with
 a set of possible worlds $\W$, the accessibility relations ${\R}_i \subseteq \W  \times
 \W$, $i = 1,2,...$, and   a mapping
 $~~I_K:P \rightarrow
\textbf{2}^{\W}$, such that for any propositional letter $p_i \in
P$, the function $~I_K(p_i):\W \rightarrow \textbf{2}~$ defines the
truth of $p_i$ in a world $w \in \W$.
 \end{definition}
 For any formula $\varphi$  we define  $~~{\M} \models_{w}~\varphi~$ iff $~\varphi$ is satisfied in a world $w \in \W$.
  For example, a given  letter $p_i$ is true in $w$,
 i.e., $~{\M} \models_{w}~p_i,~$ iff $~I_K(p_i)(w) = t$.\\
 The Kripke semantics is extended to all formulae as follows:
 \\$~~{\M} \models_{w}~ \varphi \wedge \phi~~~$ iff $~~~{\M} \models_{w}~ \varphi~$ and $~{\M} \models_{w}~
 \phi~$, \\
 $~~{\M} \models_{w}~ \neg \varphi ~~~$ iff $~~~$ not ${\M} \models_{w}~ \varphi~$, \\
  $~~{\M} \models_{w}~\lozenge_i \varphi~~~$ iff $~~~$ exists $w'\in \W$ such that $(w,w')
\in {\R}_i $ and ${\M} \models_{w'}~ \varphi$.\\
 The universal
modal operator $\square_i$ is equal to $\neg \lozenge_i \neg$.\\
A formula $\varphi$ is said to be \emph{true in a Kripke's
interpretation} ${\M}$ if  for each possible world $w$, ${\M}
\models_{w}~\varphi$. A formula is said to be \emph{valid} if it is
true in each interpretation.
 \begin{definition} \label{def:KripSem} \textsc{Predicate multi-modal
 logic}:\\
 We denote by $\M = (\W, \{$$
{\R}_i~ | ~1 \leq i \leq k \}, \D, I_K)$ a multi-modal  Kripke model
with finite $k \geq 1$ modal operators with
 a set of possible worlds $\W$, the accessibility relations ${\R}_i \subseteq \W  \times \W$,
 non empty domain $\D$, and    a mapping $~~I_K:\W\times (P \bigcup F) \rightarrow {\bigcup}_{n \in \N}
(\textbf{2}\bigcup \D)^{\D^n}$, such that for any world $w \in \W$,\\
1. For any functional letter $f_i^k \in F$, $~I_K(w,f_i^k):\D^k
\rightarrow \D~$ is a
function (interpretation of $f_i^k$ in $w$).\\
2. For any predicate letter $p_i^k \in P$, the function
$~I_K(w,p_i^k):\D^K \rightarrow \textbf{2}~$ defines the extension
of $p_i^k$ in a world $w$, \\$~~~~\|p_i^k(x_1,...,x_k)\|_{\M, w}
=_{def} \{  (d_1,...,d_k) \in \D^k~|~ ~I_K(w,p_i^k)(d_1,...,d_k) = t
\}$.
 \end{definition}
 For any formula $\varphi$  we define  $~~{\M} \models_{w,g}~\varphi~$ iff $~\varphi$ is satisfied in a world $w \in \W$ for
 a given assignment $g:\V \rightarrow \D$. For example, a given atom $p_i^k(x_1,...,x_k)$ is satisfied in $w$ by assignment $g$,
 i.e., $~{\M} \models_{w,g}~p_i^k(x_1,...,x_k),~$ iff $~I_K(w,p_i^k)(g(x_1),...,g(x_k)) = t$.\\
 The Kripke semantics is extended to all formulae as follows:
 \\$~~{\M} \models_{w,g}~ \varphi \wedge \phi~~~$ iff $~~~{\M} \models_{w,g}~ \varphi~$ and $~{\M} \models_{w,g}~
 \phi~$, \\
 $~~{\M} \models_{w,g}~ \neg \varphi ~~~$ iff $~~~$ not ${\M} \models_{w,g}~ \varphi~$, \\
  $~~{\M} \models_{w,g}~\lozenge_i \varphi~~~$ iff $~~~$ exists $w'\in \W$ such that $(w,w')
\in {\R}_i $ and ${\M} \models_{w',g}~ \varphi$.\\
A formula $\varphi$ is said to be \emph{true in a Kripke's
interpretation} ${\M}$ if  for each assignment function $g$ and
possible world $w$, ${\M} \models_{w,g}~\varphi$. A formula is said
to be \emph{valid} if it
is true in each interpretation.\\
Any virtual predicate $\phi(x_1,...,x_k)$ has different extensions
$~~~~\|\phi(x_1,...,x_k)\|_{\M, w} =_{def} \{ (g(x_1),...,g(x_k))
~|~ g \in \D^{\V}$ and $~{\M} \models_{w,g}~ \phi \}$ for different
possible worlds $w \in \W$. Thus we can not establish the simple
extensional identity for two concepts as in FOL.
 Apparently it seams that Tarski's
interpretation for the FOL and the Kripke's interpretation for modal
predicate logics are inconceivable. Currently, each modal logic is
considered as a kind of intensional logic. The open question is what
about the modality in the FOL,  if it is intrinsic also in FOL, that
is, if there is an equivalent multi-modal transformation of the FOL
where the Kripke's interpretation is an equivalent corespondent to
the original Tarski's interpretation for the FOL. The positive
answer to these questions is one of the main contributions of this
paper.
\\$\square$\\
%
%
The Plan of this work is the following: in Section 2 will be
 presented the PRP theory and the two step intensional
semantics for modal predicate logics, with the unique intensional
interpretation $I$ which maps the logic formulae into the concepts
(intensional entities), and the set of extensionalization functions
which determine the extension of any given concept in different
possible worlds. After that we will define an extensional algebra of
relations for the FOL, different from standard Cylindric algebras.
\\In Section 3 we will consider the FOL syntax with the modal
Kripke's semantics for each particular application of quantifiers
$(\exists x)$, and we will obtain a multi-modal predicate logic
FOL$_{\K}$, equivalent to the standard FOL with Tarski's
interpretation. Moreover, we will define the generalized Kripke
semantics for modal predicate logics, and we will show their diagram
of fundamental reductions, based on the restrictions over possible
worlds. In Section 4 we will consider the intensionality of modal
logics, and we will show that not all modal logics are intensional
as supposed: in fact the modal translation of the FOL syntax results
in a multi-modal predicate logic FOL$_{\K}$ that is pure extensional
as it is the standard Tarskian FOL. Then we will define the full
intensional
enrichment for multi-modal predicate logics.\\
Finally, in Section 5 we will consider the minimal intensional
enrichment of the FOL (which does not change the syntax of the FOL),
by defining FOL$_{\I}$ intensional logic with the set of explicit
possible worlds equal to the set of Tarski's interpretations of the
standard extensional FOL. We will show that its intensionality
corresponds to the Montague's point of view. Then we will define the
intensional algebra of concepts for this intensional FOL$_{\I}$, and
the homomorphic correspondence of the two-step intensional semantics
with the Tarskian semantics of the FOL, valid in every possible
world of FOL$_{\I}$.
\section{Intensionality  and intensional/extensional semantics \label{section:Intensionality}}
Contemporary use of the term 'intension' derives from the
traditional logical doctrine that an idea has both an extension and
an intension. Although there is divergence in formulation, it is
accepted that the extension of an idea  consists of the subjects to
which the idea applies, and the intension consists of the attributes
implied by the idea. In contemporary philosophy, it is linguistic
expressions (here it is a logic formula), rather than concepts, that
are said to have intensions and extensions. The intension is the
concept expressed by the expression, and the extension is the set of
items to which the expression applies. This usage resembles use of
Frege's use  of 'Bedeutung' and 'Sinn' \cite{Freg92}. It is evident
that two ideas could
have the same extension but different intensions.\\
The systematic study of intensional entities has been pursued
largely in the context of intensional logic; that part of logic in
which the principle of (extensional) substitutivity of equivalent
expressions fails.
\\
Intensional entities (or concepts) are such things as propositions,
relations and properties. What make them 'intensional' is that they
violate the principle of extensionality; the principle that
extensional equivalence implies identity. All (or most) of these
intensional entities have been classified at one time or another as
kinds of Universals \cite{Beal93}. Accordingly, standard traditional
views about the ontological status of universals carry over to
intensional entities. Nominalists hold that they do not really
exist. Conceptualist accept their existence but deem it to be
mind-dependent. Realists hold that they are mind-independent.
\emph{Ante rem} realists hold that they exist independently of being
true of anything; \emph{in re} realists require that they be true of
something \cite{Beal93}. In what follows we adopt the \emph{Ante rem} realism.\\
In a predicate logics, (virtual) predicates  expresses classes
(properties and relations), and sentences express propositions. Note
that classes (intensional entities) are \emph{reified}, i.e., they
belong to the same domain as individual objects (particulars). This
endows the intensional logics with a great deal of uniformity,
making it possible to manipulate classes and individual objects in
the same language. In particular, when viewed as an individual
object, a class can be a member of another class.\\
The extensional reductions, such as, propositional complexes and
propositional functions, to intensional entities are inadequate,
there are several technical difficulties \cite{Beal93b}, so that we
adopt the non-reductionist approaches and we will show how it
corresponds to the possible world semantics. We begin with the
informal theory that universals (properties (unary relations),
relations, and propositions in PRP theory \cite{Beal79}) are genuine
entities that bear fundamental logical relations to one another. To
study properties, relations and propositions, one defines a family
of set-theoretical structures, one define the intensional algebra, a
family of set-theoretical structures most of which are built up from
arbitrary objects and fundamental logical operations (conjunction,
negation,
existential generalization,etc..) on them.\\
The value of both traditional conceptions of PRPs (the 'possible
worlds' and 'algebraic' Russel's approaches) is evident, and in
Bealer's work both conceptions are developed side by side
\cite{Beal82}. But Bealer's approach to intensional logic locates
the origin of intensionality a single underlying \emph{intensional
abstraction} operation which transforms the logic formulae into
terms, so that we are able to make reification of logic formulae
without the necessity of the second-order logics. In fact, the
\emph{intensional abstracts} are so called 'that'-clauses. We assume
that they are singular terms; Intensional expressions like
'believe', mean', 'assert', 'know',
 are standard two-place predicates  that take 'that'-clauses as
 arguments. Expressions like 'is necessary', 'is true', and 'is
 possible' are one-place predicates that take 'that'-clauses as
 arguments. For example, in the intensional sentence "it is
 necessary that A", where $A$ is a proposition, the 'that A' is
 denoted by the $\langle A \rangle$, where $\langle \rangle$ is the intensional abstraction
 operator which transforms a logic formula $A$ into the \emph{term} $\langle A \rangle$. So
 that the sentence "it is
 necessary that A" is expressed by the logic atom $N(\langle A \rangle)$, where
 $N$ is the unary predicate 'is necessary'. In this way we are able
 to avoid to have the higher-order syntax for our \emph{intensional} logic language
 (predicates appear in variable places of other predicates),as, for example HiLog \cite{ChKW93} where the \emph{same}
 symbol may denote a predicate, a function, or an atomic formula. In
 the First-order logic (FOL) with intensional abstraction we have
 more fine distinction between an atom $A$ and its use as a
 \emph{term} 'that A', denoted by $\langle A \rangle$ and considered
 as intensional 'name',  inside some other predicate, and, for example, to have the first-order formula $\neg A \wedge P(t, \langle A
 \rangle)$ instead of the second-order HiLog formula $\neg A \wedge P(t,  A
 )$. \\
In this work I will not accept this Baler's approach, and I will
consider the \emph{minimal} intensionality in FOL \emph{without}
necessity of intensional abstraction operation. Thus I will consider
only basic conceptions of intensional entities: open-sentences
(transformed into virtual predicates with non empty tuple of free
variables) express properties and relations, and sentences express
propositions. But the concepts (properties, relations and
propositions) are denotations for open and closed logic sentences,
thus elements of the structured domain  $~\D = D_{-1} + D_I$, (here
$+$ is a disjoint union) where a subdomain $D_{-1}$ is made of
 particulars (individuals), and the rest $D_I = D_0 +
 D_1 ...+ D_n ...$ is made of
 universals (\emph{concepts}): $D_0$ for  propositions with a distinct element $Truth \in D_0$, $D_1$ for properties
 (unary concepts)
  and  $D_n, n \geq 2,$ for n-ary concept.
  The
 concepts in $\D_I$ are denoted by $u,v,...$, while the
 values (individuals) in $D_{-1}$ by $a,b,...$ (the empty tuple $<>$ of the nullary relation is an individual in
 $D_{-1}$, with $\D^0 = \{<>\}$,
 so that $\{f,t\} = \P(\D^0) \subseteq \P(D_{-1})$).\\
 Sort $S$ is a subset of a domain $\D$. For example  $[0,1]$ is
 closed-interval of reals sort, $\{0,1,2,3,..\} \subseteq D_{-1}$ is the sort of integers, etc..
 These sorts are used for sorted variables in many-sorted
 predicate logics so that the assigned values for each sorted variable must belong to its sort.
  The unsorted variables can be considered as variables with a
 top sort equal to $\D$. \\
 The \emph{intensional interpretation} is a mapping between the set $\L$ of formulae of the logic language  and
 intensional entities in $\D$, $I:\L \rightarrow \D$, is a kind of
 "conceptualization", such that  an open-sentence (virtual
 predicate)
 $\phi(x_1,...,x_k)$ with a tuple of all free variables
 $(x_1,...,x_k)$ is mapped into a k-ary \emph{concept}, that is, an intensional entity  $u =
 I(\phi(x_1,...,x_k)) \in D_k$, and (closed) sentence $\psi$ into a proposition (i.e., \emph{logic} concept) $v =
 I(\psi) \in D_0$ with $I(\top) = Truth \in D_0$ for the FOL tautology $\top$. A language constant $c$ is mapped into a
 particular $a = I(c) \in D_{-1}$ if it is a proper name, otherwise in a correspondent concept in $\D$.
\begin{definition} \label{def:extent} \textsc{Extensions and extensionalization functions:}\\
 Let $\mathfrak{R} = \bigcup_{k \in \mathbb{N}} \P(\D^k) = \sum_{k\in \mathbb{N}}\P(D^k)$ be the set of all k-ary relations, where $k \in \mathbb{N} =
\{0,1,2,...\}$. Notice that $\{f,t\} = \P(\D^0) \in \mathfrak{R}$,
that is, the truth values are extensions in $\mathfrak{R}$.
 The extensions of the intensional entities (concepts) are given by the set
 $\E$ of extensionalization functions $h:\D \rightarrow
 \mathfrak{R}$, such that
 \begin{center}
 $h = h_{-1} + h_0 + \sum_{i\geq 1}h_i:\sum_{i
\geq -1}D_i \longrightarrow \P(D_{-1}) + \{f,t\} + \sum_{i\geq
1}\P(D^i)$
\end{center}
 where $h_{-1} = id:D_{-1} \rightarrow D_{-1}$
is an identity,
$~h_0:D_0 \rightarrow \{f,t\} = \P(\D^0)$ assigns the truth values
in $ \{f,t\}$ to all propositions with the constant assignment
$h_0(Truth) = t$, and $h_i:D_i \rightarrow \P(D^i)$, $i\geq 1$,
assigns an extension to each
concept.\\
Consequently, intensions can be seen as \verb"names" of abstract or
concrete concepts, while extensions correspond to various rules that
these concepts play in different worlds.
\end{definition}
 Thus, for any open-sentence $\phi(x_1,...,x_k)$ we have that
its extension, in a given world $w \in \W$ of the Kripke's
interpretation $\M = (\W, \{{\R}_i~ | ~1 \leq i \leq k \}, \D, V)$
for modal (intensional) logics in Definition \ref{def:KripSem}, is
equal to: $~~~~h(I(\phi(x_1,...,x_k))) = \\ =
\|\phi(x_1,...,x_k)\|_{\M, w} =
 \{ (g(x_1),...,g(x_k)) ~|~ g \in \D^{\V}$ and $~{\M}
\models_{w,g}~ \phi \}$.\\
From a logic point of view, two possible worlds $w$ and $w'$ are
indistinguishable if all sentences have the same extensions in them,
so that we can consider an extensionalization function $h$ as a
"possible world", similarly to the semantics of a probabilistic
logic, where possible worlds are Herbrand interpretations for given
set of predicate letters $P$ in a given logic. Thus, for a given
modal logic we will have that there is a mapping $is:\W \rightarrow
\E$ from the set of possible worlds to the set of
 extensionalization functions.
 \begin{definition} \label{def:intensemant} \textsc{Two-step \textsc{I}ntensional \textsc{S}emantics:}
 The
intensional semantics of the logic language with the set of formulae
$\L$ can be represented by the  mapping
\begin{center}
$~~~ \L ~\longrightarrow_I~ \D ~\Longrightarrow_{w \in \W}~
\mathfrak{R}$,
\end{center}
where $~\longrightarrow_I~$ is a \emph{fixed intensional}
interpretation $I:\L \rightarrow \D$ and $~\Longrightarrow_{w \in
\W}~$ is \emph{the set} of all extensionalization functions $h =
is(w):\D \rightarrow \mathfrak{R}$ in $\E$, where $is:\W \rightarrow
\E$ is the mapping from the set of possible worlds to the set of
 extensionalization functions.\\
 We define the mapping $I_n:\L_{op} \rightarrow
\mathfrak{R}^{\W}$, where $\L_{op}$ is the subset of formulae with
free variables (virtual predicates), such that for any virtual
predicate $\phi(x_1,...,x_k) \in \L_{op}$ the mapping
$I_n(\phi(x_1,...,x_k)):\W \rightarrow \mathfrak{R}$ is the
Montague's meaning (i.e., \emph{intension}) of this virtual
predicate \cite{Lewi86,Stal84,Mont70,Mont73,Mont74}, that is, the
mapping which returns with the extension of this (virtual) predicate
in every possible world in $\W$.
\end{definition}
We adopted this two-step intensional semantics, instead of well
known Montague's semantics (which lies in the construction of a
compositional and recursive semantics that covers both intension and
extension) because of a number of its weakness. Let us consider the
following two past participles: 'bought' and 'sold'(with unary
predicates $p_1^1(x)$, '$x$ has been bought', and $p_2^1(x)$,'$x$
has been sold'). These two different concepts in the Montague's
semantics would have not only the same extension but also their
intension, from the fact that their extensions are identical in
every possible world. Within the two-steps formalism we can avoid
this problem by assigning two different concepts (meanings) $u =
I(p_1^1(x))$ and $v = I(p_2^1(x))$ in $ D_1$. Notice that the same
problem we have in the Montague's semantics for two sentences with
different meanings, which bear the same truth value across all
possible worlds: in the Montague's semantics they will be forced to
the \emph{same} meaning.\\
But there is also another advantage of this two-step intensional
semantics in Definition \ref{def:intensemant}: here we are able to
define an intensional algebra $\A_{int}$ over intensional entities
in $\D$, which is autosufficient, differently from Montague's
semantics where the compositional and recursive semantics of
intensions can be defined only by their extensional properties. As
we will see in the last Section, this intensional algebra is defined
in the way that each extensional mapping $h =is(w):\D \rightarrow
\mathfrak{R}$
 is a \emph{homomorphism} between this intentensional algebra $\A_{int}$ and the extensional relational algebra $\A_{\mathfrak{R}}$ that represents
 the compositional and recursive semantics of the extensions,
 given by Corollary \ref{coro:intensemant} later in this Section. In this way the compositional and recursive semantics of the intensions in $\A_{int}$
 coincide with the Montague's semantics, where, for example, the mapping $I_n(\phi\wedge \psi):\W \rightarrow \mathfrak{R}$ (i.e., the Montague's intension of the
 composite formula $\phi\wedge \psi$) is
 functionally dependent on the mappings $I_n(\phi):\W \rightarrow \mathfrak{R}$ and  $I_n(\psi):\W \rightarrow \mathfrak{R}$ (i.e.,
  dependent on the Montague's intensions of $\phi$ and $\psi$).\\
  \textbf{Remark:} the mapping $I_n$ can be
extended also to all sentences (the formulae without free
variables), such that for any sentence $\phi$, $I_n(\phi):\W
\rightarrow \{f,t\} = \P(\D^0) \subseteq \mathfrak{R}$ is a mapping
that defines the truth value (i.e., an extension in $\mathfrak{R}$
in Definition \ref{def:extent}) of this sentence in each possible
world $\W$. Equivalently to this Montague's semantics for intensions
of logic formulae, we can use the Carnap's semantics \cite{Carn47}
of concepts in $\D$, that is $I_{n,c}:\D \rightarrow
\mathfrak{R}^{\W}$ such hat the intension of a \emph{concept} $u \in
\D$ is a mapping $I_{n,c}(u):\W \rightarrow \mathfrak{R}$ from
possible worlds to extensions. This Carnap's semantics
of concepts is represented by the second mapping of the diagram $\D ~\Rightarrow_{w \in \W}~ \mathfrak{R}$ above.\\
 Tarski's interpretation of the
FOL  is instead given by a single mapping $I_T^*:\L\rightarrow
\mathfrak{R}$, as explained in the introduction dedicated to FOL.
Thus, if there is a modal Kripke semantics with a set of possible
worlds $\W$ (thus, an intensional semantics) for FOL,
\emph{equivalent} to the standard FOL semantics given by the
Tarski's interpretation $I_T$, then we have to obtain for every
possible world $w \in \W$ of such a semantics that $h = is(w)$ is
invariant (i.e., the set $\{h = is(w)~|~w \in \W\}$ is a singleton
set), and consequently $I_T^* = h \circ I$, where $\circ$ is a
composition of functions, such that for any formula $\phi \in \L$,
$h(I(\phi)) = I_T^*(\phi)$. For any constant $c$ of the FOL language
 we assume that $I(c) = I_T(c) \in \D$.
\\$\square$\\
We consider that the domain $\D$ is equal in each possible world $w
\in \W$. It is demonstrated that also in the case of different
domains $\D_w$ in different possible worlds, we can always obtain
the constant domain model (as in Definition 2.1 in \cite{Fitt01})
$\D = \bigcup_{w \in \W} \D_w$ and by introducing a new built-in
binary predicate $e(x,y)$ where $x$ has as domain the set of
possible worlds, so that $e(w,d)$ is true if $d \in \D_w$. It is
important that the set of particulars $\D_{-1}$ is the set of
\emph{rigid objects} like "Eiffel tower" or "George Washington",
that have equal extension (denotation) in each possible world: it
holds from the fact that for every rigid object $c$, a possible
world $w \in \W$, and a given intensional interpretation $I$ we have
that $d = I(c) \in \D_{-1}$ and its extension is $h(d)= is(w)(d) =
d$ constant independently from $w$.\\
The problem of \emph{non-rigid objects} and relative complications
considered by Fitting in \cite{FiMe98,Fitt01,Fitt04}, as "the gross
domestic product of Denmark" or "the Secretary-General of the United
Nations", here are considered not as constants of the language but
as unary predicates, denoted by $p_1^1, p_2^1 \in P$. The intension
$ u = I(p_1^1) \in D_1$  denotes the property (unary concept) whose
extension $is(w)(I(p_1^1))$ is a singleton set (by introducing an
axiom $(\exists_1 x)p_1^1(x)$), possibly different in each possible
world (for example, the instance of time) $w \in \W$. If we need to
use these "non-rigid objects" as arguments inside other predicates,
in order to avoid
 the second-order syntax we can use Bealer's \emph{intensional abstraction} \cite{Beal82} which can transform the unary predicates used for non-rigid objects into terms, so that
 can be used as arguments inside other predicates. \\
It explains why in these two-step interpretations, intensional and
extensional, we can work in an unified \emph{general rigid}
framework, and overcome the major difficulties for modal first-order
logics, considered by Fitting in the number of his papers, by
introducing new operations like 'extension of' operators
$\downarrow$ and 'predicate abstracts', $<\lambda
x_1,...,\\x_n.\phi>$ that transforms the logic formula with a tuple
of free variables $\phi(x_1,...,x_n)$ into new  \emph{atomic}
formula $<\lambda x_1,...,x_n.\phi>(t_1,...,t_n)$ for any given set
of terms $t_i, i = 1,...,n$ (Definition 2.3 in \cite{Fitt01}).
Notice that differently from this Fitting's approach we do not
consider a virtual predicate $\phi(x_1,...,x_n)$, as a new
\emph{atom}, but as
a standard logic formula.\\
Another relevant question w.r.t. this two-step interpretations of an
intensional semantics is how in it is managed  the extensional
identity relation $\doteq$ (binary predicate of the identity) of the
FOL. Here this extensional identity relation is mapped into the
binary concept $Id = I(\doteq(x,y)) \in D_2$, such that $(\forall w
\in \W)(is(w)(Id) = R_{=})$,
where $\doteq(x,y)$ denotes an atom of the FOL
of the binary predicate for identity in FOL, usually written by FOL
formula $x \doteq y$ (here we prefer to distinguish this
\emph{formal symbol} $~ \doteq ~ \in P$ of the built-in identity
binary predicate letter in the FOL from the standard mathematical
symbol '$=$' used in all mathematical definitions in this paper).
That is, for every possible world $w$ and its correspondent
extensionalization function $h = is(w)$, the extensional identity
relation in $\D$ is the extension of the binary concept  $Id \in
D_2$, as defined by Bealer's approach to intensional FOL with
intensional abstraction in
\cite{Beal82}.\\
Let $\A_{FOL} = (\L, \doteq, \top, \wedge, \neg, \exists)$ be a free
syntax algebra for "First-order logic with identity $\doteq$", with
the set $\L$ of first-order logic formulae,  with $\top$ denoting
the tautology formula (the contradiction formula is denoted by $
\neg \top$), with the set of variables in $\V$ and the domain of
values in $\D$ . It is well known that we are able to make the
extensional algebraization of the FOL by using the \emph{cylindric}
algebras \cite{HeMT71} that are the extension of Boolean algebras
with a set of binary operators for the FOL identity relations and a
set of unary algebraic operators ("projections") for each case of
FOL quantification $(\exists x)$. In what follows we will make an
analog extensional algebraization over $\mathfrak{R}$ but by
interpretation of the logic conjunction $\wedge$ by a set of
\emph{natural join} operators over relations introduced by Codd's
relational algebra \cite{Codd70,Piro82} as a kind of a predicate
calculus whose interpretations are tied to the
database.\\
 In what follows we will use the function $f_{<>}:\mathfrak{R}
\rightarrow \mathfrak{R}$, such that for any $R \in \mathfrak{R}$,
$f_{<>}(R) = \{<>\}$ if $R \neq \emptyset$; $\emptyset$ otherwise.
Let us define the following set of algebraic operators for
 relations in $\mathfrak{R}$:
\begin{enumerate}
\item binary operator $~\bowtie_{S}:\mathfrak{R} \times \mathfrak{R} \rightarrow
\mathfrak{R}$,
 such that for any two relations $R_1, R_2 \in
 \mathfrak{R}~$, the
 $~R_1 \bowtie_{S} R_2$ is equal
to the relation obtained by natural join
 of these two relations $~$ \verb"if"
 $S$ is a non empty
set of pairs of joined columns of respective relations (where the
first argument is the column index of the relation $R_1$ while the
second argument is the column index of the joined column of the
relation $R_2$); \verb"otherwise" it is equal to the cartesian
product $R_1\times R_2$.\\ For example, the logic formula
$\phi(x_i,x_j,x_k,x_l,x_m) \wedge \psi (x_l,y_i,x_j,y_j)$ will be
traduced by the algebraic expression $~R_1 \bowtie_{S}R_2$ where
$R_1 \in \P(\D^5), R_2\in \P(\D^4)$ are the extensions for a given
Tarski's interpretation $I_T$ of the virtual predicate $\phi, \psi$
relatively, so that $S = \{(4,1),(2,3)\}$ and the resulting relation
will have the following ordering of attributes:
$(x_i,x_j,x_k,x_l,x_m,y_i,y_j)$.
 Consequently, we have that for
any two formulae $\phi,\psi \in \L$ and a particular join operator
$\bowtie_{S}$ uniquely determined by tuples of free variables in
these two formulae,\\ $~I_T^*(\phi \wedge \psi) = I_T^*(\phi)
\bowtie_{S} I_T^*(\psi)$.
\item unary operator $~ \sim:\mathfrak{R} \rightarrow \mathfrak{R}$, such that for any k-ary (with $k \geq 0$)
relation $R \in  \P(\D^{k}) \subset \mathfrak{R}$
 we have that $~ \sim(R) = \D^k \backslash R \in \D^{k}$, where '$\backslash$' is the substraction of relations.\\ For example, the
logic formula $\neg \phi(x_i,x_j,x_k,x_l,x_m)$ will be traduced by
the algebraic expression $~\D^5 \backslash R$ where $R$ is the
extensions for a given Tarski's interpretation $I_T$ of the virtual
predicate $\phi$.
Consequently, we have that for any  formula $\phi \in \L$,\\
$I_T^*(\neg \phi) = \sim( I_T^*(\phi))$.
\item unary operator $~ \pi_{-m}:\mathfrak{R} \rightarrow \mathfrak{R}$, such that for any k-ary (with $k \geq 0$) relation $R \in \P(\D^{k}) \subset \mathfrak{R}$
we have that $~ \pi_{-m} (R)$ is equal to the relation obtained by
elimination of the m-th column of the relation $R~$ \verb"if" $1\leq
m \leq k$ and $k \geq 2$; equal to $~f_{<>}(R)~$ \verb"if" $m = k
=1$; \verb"otherwise" it is equal to $R$.\\ For example, the logic
formula $(\exists x_k) \phi(x_i,x_j,x_k,x_l,x_m)$ will be traduced
by the algebraic expression $~\pi_{-3}(R)$ where $R$ is the
extensions for a given Tarski's interpretation $I_T$ of the virtual
predicate $\phi$ and the resulting relation will have the following
ordering of attributes: $(x_i,x_j,x_l,x_m)$.  Consequently, we have
that for any formula $\phi \in \L$ with a free variable $x$,  where
$m$ is equal to the position of this
 variable $x$ in the tuple of free variables in $\phi$ (or $m = 0$ otherwise, where $\pi_{-0}$ is the identity function),
$~I_T^*((\exists x)\phi) = \pi_{-m}(I_T^*(\phi))$.
\end{enumerate}
Notice that the ordering of attributes of resulting relations
corresponds to the method used for generating the ordering of
variables in the tuples of free variables adopted for virtual
predicates, as explained in the introduction to FOL.
 \begin{coro} \label{coro:intensemant} \textsc{Extensional FOL
 semantics:}\\
Let us define the extensional relational algebra  for the FOL by,\\
$\A_{\mathfrak{R}} = (\mathfrak{R}, R_=, \{<>\}, \{\bowtie_{S}\}_{ S
\in \P(\mathbb{N}^2)}, \sim, \{\pi_{-n}\}_{n \in \mathbb{N}})$,
\\where $ \{<>\} \in \mathfrak{R}$ is the algebraic value
correspondent to the logic truth, and $R_=$ is the binary relation
for extensionally equal elements.
We will use '$=$' for the extensional identity for relations in $\mathfrak{R}$.\\
Then, for any Tarski's interpretation $I_T$ its unique extension to
all formulae $I_T^*:\L \rightarrow \mathfrak{R}$ is also the
homomorphism $I_T^*:\A_{FOL} \rightarrow \A_{\mathfrak{R}}$ from the
free syntax FOL algebra into this extensional relational algebra.
\end{coro}
\textbf{Proof:} Directly from definition of the semantics of the
operators in $\A_{\mathfrak{R}}$ defined in precedence. Let us take
the case of conjunction of logic formulae of the definition above
where $\varphi(x_i,x_j,x_k,x_l,x_m,y_i,y_j)$ (it's tuple of
variables is obtained by the method defined in the FOL introduction)
is the virtual predicate of the logic formula
$\phi(x_i,x_j,x_k,x_l,\\x_m) \wedge \psi (x_l,y_i,x_j,y_j)$:
$~~I_T^*(\phi \wedge \psi) = \\ = I_T^*(\varphi) =
\{(g(x_i),g(x_j),g(x_k),g(x_l),g(x_m),g(y_i),g(y_j))~|~I_T^*(\varphi/g)
= t\}\\ =
\{(g(x_i),g(x_j),g(x_k),g(x_l),g(x_m),g(y_i),g(y_j))~|~I_T^*(\phi/g
\wedge \psi/g) = t\}\\ =
\{(g(x_i),g(x_j),g(x_k),g(x_l),g(x_m),g(y_i),g(y_j))~|~I_T^*(\phi/g)
= t$ and $I_T^*(\phi/g) = t\}\\ =
\{(g(x_i),g(x_j),g(x_k),g(x_l),g(x_m),g(y_i),g(y_j))~|~(g(x_i),g(x_j),g(x_k),g(x_l),g(x_m))
\\ \in I_T^*(\phi)$ and $(g(x_l),g(y_i),g(x_j),g(y_j))\in I_T^*(\phi)\}
\\= I_T^*(\phi) \bowtie_{\{(4,1),(2,3)\}} I_T^*(\psi)$.\\
 Thus, it is enough to show that is valid also $I_T^*(\top) =
\{<>\}$, and
 $I_T^*(\neg \top) = \emptyset$. The first
property comes from the fact that $\top$ is a tautology, thus
satisfied by every assignment $g$, that is it is true, i.e.
$I_T^*(\top) = t$ (and $t$ is equal to the empty tuple $\{<>\}$).
The second property comes from the fact that $I_T^*(\neg\top) =
 \sim(I_T^*(\top)) = \sim(
\{<>\})\\ = \D^0 \backslash \{<>\} = \{<>\}\backslash \{<>\} =
\emptyset$.  That is, the tautology and the contradiction have the
true and false logic value respectively in $\mathfrak{R}$.\\ We have
also that $I_T^*(\doteq(x,y)) = I_T(\doteq) = R_=$ for every
interpretation $I_T$ because $\doteq$ is the built-in binary
predicate, that is, with the same extension in every Tarski's
interpretation.
\\Consequently, the mapping $I_T^*:(\L, \doteq, \top,
\wedge, \neg, \exists) \rightarrow \A_{\mathfrak{R}}$ is a
homomorphism that  represents the extensional Tarskian semantics of
the FOL.
\\$\square$\\
Notice that $\mathfrak{R}$ is a poset with the bottom element
$\emptyset$ and the top element $\{<>\}$, and  the partial ordering
$\preceq$ defined as follows: for any two relations $R_1,
R_2 \in \mathfrak{R}$, \\
$R_1 \preceq R_2$ iff "for some operation $\bowtie_{S}$ it
holds that $(R_1 \bowtie_{S} R_2) = R_1$".\\
It is easy to verify that for any $R \in \mathfrak{R}$ and operation
$\bowtie_{S}$ it holds that $(R \bowtie_{S} \emptyset) = \emptyset$,
and $(R \bowtie_{S} \{<>\}) = R$. That is, $\emptyset \preceq R
\preceq \{<>\}$.
\section{First-order logic and  modality \label{section:FOLmod} }
In propositional modal logics the possible worlds are entities where
a given propositional symbol can be true or false. Thus, from
\emph{logical} point of view the \emph{possible worlds}
\cite{Wans90,WaPe89} in Kripke's relational semantics are
characterized by property to determine the truth of logic sentences.
The important question relative to the syntax of the FOL is  if
there is a kind of basic set of possible worlds that have such
properties. The answer is
affirmative.\\
In fact, if we consider a k-ary predicate letter $p_i^k$ as a new
kind of 'propositional letter', then an assignment $g:\V \rightarrow
\D$ can be considered as an \emph{intrinsic} (par excellence)
possible world, where the truth of this 'propositional letter'
$p_i^k$ is equal to the truth of the ground atom
$p_i^k(g(x_1),...,g(x_k))$.\\
 Consequently, in what follows we will
denote by $\W$ the set of \emph{explicit} possible worlds (defined
explicitly for each particular case of
 modal logics), while the set $\D^{\V}$ we will be called as the set of
\emph{intrinsic} possible worlds (which is invariant and common for
every \emph{predicate} modal logic). In the case when $\V =
\emptyset$ is the empty set we obtain  the singleton set of
intrinsic possible worlds $\D^{\V} =
\{*\}$, with the empty function $*:\emptyset \rightarrow \D$. \\
By $\mathbb{W} \subseteq \{(w,g)~|~ w \in \W, g \in \D^{\V}\}$ we
will denote the set of (generalized) possible worlds. In this way,
as in the case of propositional modal logic, we will have that a
formula $\varphi$ is \emph{true} in a Kripke's interpretation ${\M}$
if  for each (generalized) possible world $u = (w,g) \in
\mathbb{W}$, $~~{\M} \models_{u}~\varphi$.\\
We denote by $|\phi| = \{(w,g) \in \mathbb{W}~|~\M \}
\models_{w,g}~\phi\}$ the set of all worlds where the  formula
$\phi$ is satisfied by interpretation $\M$. Thus, as in the case of
propositional modal logics, also in the case of predicate modal
logics
we have that a formula $\phi$ is true iff  it is satisfied in all (generalized) possible worlds, i.e., iff $|\phi| = \mathbb{W}$.\\
With this new arrangement we can reformulate the standard semantics
for multimodal predicate logic in Definition \ref{def:KripSem}(
$\pi_1$ and $\pi_2$ denote the first and the second projections):
\begin{definition} \label{def:NewKripSem} \textsc{Generalized Kripke semantics
for multi-modal logics}:\\
 We denote by $\M = (\mathbb{W}, \{$$
{\R}_i \}, \D, I_K)$ a multi-modal  Kripke's interpretation with
 a set of (generalized) possible worlds $\mathbb{W}$, a set of explicit possible worlds $\W = \pi_1(\mathbb{W})$ and $ \pi_2(\mathbb{W}) = \D^{\V}$, the accessibility relations ${\R}_i \subseteq \W \times
 \W$,  $i = 1,2,...$,
 non empty domain $\D$,
and    a mapping $~~I_K:\W\times (P \bigcup F) \rightarrow
{\bigcup}_{n \in \N}
(\textbf{2}\bigcup \D)^{\D^n}$, such that for any world $w \in \W$,\\
1. For any functional letter $f_i^k \in F$, $~I_K(w,f_i^k):\D^k
\rightarrow \D~$ is a
function (interpretation of $f_i^k$ in $w$).\\
2. For any predicate letter $p_i^k \in P$, the function
$~I_K(w,p_i^k):\D^K \rightarrow \textbf{2}~$ defines the extension
of $p_i^k$ in a world $w$, \\$~~~~\|p_i^k(x_1,...,x_k)\|_{\M, w}
=_{def} \{  (d_1,...,d_k) \in \D^k~|~ ~I_K(w,p_i^k)(d_1,...,d_k) = t
\}$.
 \end{definition}
 Now we will see that we have two particular "projections" of the
 generalized Kripke semantics for multi-modal logics, defined above:
 the first one is explicit-worlds "projection" resulting in
 the Kripke semantics of multi-modal propositional logics; the second
  one is intrinsic-worlds "projection" resulting in the Kripke semantics of FOL
  logic.
\begin{propo} \textsc{Explicit-worlds "projection" of generalized semantics:}\\
The Kripke semantics of multi-modal propositional logic given by
Definition \ref{def:KripSemProp} is a particular case of the
Definition \ref{def:NewKripSem} when $\D, F, \V$ are empty sets and
$P$ has only nullary symbols, that is, the propositional symbols.
\end{propo}
\textbf{Proof:} In this case when $\V = \emptyset$ is the empty set
we have that $\D^{\V}$ is a singleton set, denoted by $\{*\}$, with
unique element equal to the empty function $*:\V \rightarrow \D$
(i.e., the function whose graph is empty). Thus $\mathbb{W} = \W
\times \{*\}$ is equivalent to the set of explicit worlds $\W$, so
that the original satisfaction relation $~{\M} \models_{w,g}$, where
$g = *$, of predicate modal logic in Definition \ref{def:KripSem}
can be equivalently reduced to the satisfaction relation $~{\M}
\models_{w}$ for only explicit worlds of propositional logic in
Definition \ref{def:KripSemProp}. \\
While  $~I_K:\W\times P \rightarrow {\bigcup}_{n \in \N}
\textbf{2}^{\D^n}$ where in this case  $\N = \{0\}$ we obtain is
reduced to $~I_K:\W\times P \rightarrow  \textbf{2}^{\D^0}$, where
$\D^0 = \{<>\}$ is a singleton set, thus $\textbf{2}^{\D^0}$ is
equivalent to $\textbf{2}$, so that we obtain the reduction into the
mapping $~I_K:\W\times P \rightarrow  \textbf{2}$, and by currying
(the $\lambda$ abstraction), we obtain the mapping $I'_K:P
\rightarrow \textbf{2}^{W}$, such that for any $p_i \in P$ and $w
\in \W$ we obtain that $I_K(w,p_i) = I'_K(p_i)(w) \in \textbf{2}$ is
the truth value of propositional letter (nullary predicate symbol in
$P$) in the explicit possible world $w$. It is easy to verify that
this obtained mapping $I'_K$ is that of the \emph{propositional}
modal logic given in Definition \ref{def:KripSemProp}.
\\$\square$\\
\textbf{Remark:} an interesting consequence of this explicit-world
"projection" of the generalized Kripke semantics is the idea of the
\emph{extension} of a propositional nullary predicate symbol
$p_i^0$, denoted here as a propositional symbol $p_i$, given by
Definition \ref{def:NewKripSem} by $~\|p_i\|_{\M, w} =
\|p_i^0\|_{\M, w} =_{def} \{ (d_0) \in \D^0 = \{<>\}~|~
~I_K(w,p_i^0) = t \} = \{ <>~|~ ~I'_K(p_i)(w) = t \} =
I'_K(p_i)(w)$. That is, it is equal to $t = \{<>\}$ if $p_i$ is true
in the explicit world $w$, or equal to $f$ (the empty set) if $p_i$
is false in the explicit world $w$. It is analogous to the
consideration of extensions of sentences defined in the intensional
semantics, as defined in  Section \ref{section:Intensionality},
where the \emph{truth} is the extension of sentences, distinct from
their \emph{meaning} that is, their intension (from  Montague's
point of view, the intension of the propositional letter $p_i$ used
above would be the function
$I'_K(p_i):\W \rightarrow \textbf{2}$).\\
 It is well known for a predicate logic and FOL (which is a
predicate logic extended by logic quantifiers) that we have not any
defined set of \emph{explicit} possible worlds. Thus, trying to
define the Kripke semantics (given by Definition
\ref{def:NewKripSem}) to FOL, we can  assume that the generalized
possible worlds coincide with the intrinsic possible worlds, that
is, $\mathbb{W} = \D^{\V}$.
In the case of the pure predicate logic (without quantifiers) we do
not need the possible worlds:  a ground atom in order to be true in
such a Kripke's interpretation has to be true in \emph{every}
possible world (or, alternatively, it has to be false in every
possible world). Consequently, in the predicate logics the truth of
ground atoms and sentences is invariant w.r.t. the possible worlds,
which renders unuseful the definition of possible worlds and
Kripke's semantics for these logics. But in the case of the modal
interpretation of the FOL,  the FOL quantifiers  has to be
interpreted by modal operators and their accessibility relations:
thus, the possible worlds are necessary in order to determine the
truth of logic formulae with quantifiers.\\
 Differently from the FOL with original Tarski's interpretation
for the unique existential operator $\exists$, the modal point of
view for the FOL with Kripke's interpretation have a particular
existential modal operator $\diamondsuit_x$, here denoted by
$(\exists x)$, for each variable $x \in \V$.
As usual, the universal modal
operators are defined by $(\forall x) = \neg (\exists x) \neg$.\\
Consequently the \emph{same syntax} for the FOL of a formula
$(\exists x) \phi$ can have \emph{two} equivalent semantics: the
original Tarski's interpretation that interprets the unique
existential operator $\exists$ for a variable $x$ and parenthesis
$(,)$, and Kripke's relational interpretation where the whole
expression $(\exists x)$ is interpreted as \emph{one particular}
existential modal operator $\diamondsuit_x$. This is  valid approach
based on the fact that, from the algebraic point of view, the syntax
of $(\exists x)$ can be interpreted as an unary operation which is
\emph{additive}, that is, it holds that $(\exists x)(\phi \vee \psi)
= (\exists x)(\phi) \vee (\exists x)(\psi)$, and \emph{normal},
i.e., $(\exists x)(\bot) = \bot$ where $\bot$ denotes a
contradiction sentence (the negation of the tautology $\top$); this
property is common for all \emph{existential modal} operators of the
\emph{normal} Kripke modal logics. In fact, the generalization
inference rule (G) of FOL here becomes the rule of necessitation,
and the axiom (A1) a particular case of Kripke axiom of normal modal
logics.
\\
 Thus, we have the following
particular case of Definition \ref{def:NewKripSem} (here the symbol
"$\backslash$" is the set substraction operation) :
%
\begin{definition} \label{def:FOL-KripSem} \textsc{Intrinsic-worlds "projection" of generalized
semantics:}\\
 We denote by   $\M = (\mathbb{W}, \{
{\R}_x~ | ~x \in \V \}, \D, I_K)$ a multimodal Kripke's
interpretation of the FOL,   with
 a set of (generalized) possible worlds   $ \mathbb{W} = \D^{\V}$,
  equal to the set of intrinsic possible worlds (assignments)  $ \D^{\V}$,
 the accessibility relation
 ${\R}_x = \{(w_1,w_2) \in \mathbb{W}^2~ | ~x\in \V$ and for all $y \in \V \backslash \{x\}(w_1(y) = w_2(y))\}$
 for existential modal operator $(\exists x)$ for each variable $x \in
\V$,
 non empty domain $\D$, and   a mapping $~~I_K:\mathbb{W}\times (P \bigcup F) \rightarrow
{\bigcup}_{n \in \N}
(\textbf{2}\bigcup \D)^{\D^n}$, such that for any world $w \in \mathbb{W}$,\\
1. For any functional letter $f_i^k \in F$, $~I_K(w,f_i^k):\D^k
\rightarrow \D~$ is a function (interpretation of $f_i^k$ in $w$).\\
2. For any predicate letter $p_i^k \in P$, the function
$~I_K(w,p_i^k):\D^K \rightarrow \textbf{2}~$ defines the extension
of $p_i^k$ in a world $w$.\\
Such an interpretation is the Kripke \verb"model" of the FOL  if,
for any $~ (d_1,...,d_k) \in \D^k$, for all $~w' \in \mathbb{W}
~(I_K(w',p_i^k)(d_1,...,d_k) = I_K(w,p_i^k)(d_1,...,d_k)$ and
$~I_K(w',f_i^k)(d_1,...,d_k)\\ = I_K(w,f_i^k)(d_1,...,d_k))$.\\
 \end{definition}
 We will denote by
 "FOL$_{\K}$" the FOL with
 these modal Kripke models. We recall that  FOL$_{\K}$ has \emph{the same syntax} as FOL, that is, the same set of
 formulae, and the same domain $\D$,
 differently from the standard embedding of the modal predicate logics into the FOL:
 it introduces a new built-in predicate symbol for each binary accessibility relation $\R_i$, and
 enlarges the original domain $\D$ with the set of possible worlds, and  enlarges the set
  of variables $\V$ by the new variables for these new built-in
  symbols.\\
 It is easy to verify that each accessibility binary relation
 $\R_x$, $x \in \V$, is  reflexive, transitive and symmetric
 relation. Thus, each pair of modal operators $(\exists x)$ and $(\forall x)$
 is an example of  existential and universal modal operators of the
 S5 modal logics, so that $(\forall x)$ is an "\emph{it is known} for all values assigned to  $x$ that"
 modal operator, whose semantics is equivalent to standard FOL "for all values assigned to
 $x$" semantics.\\
  It is analogous to monadic algebras of Halmos \cite{Halm62} and his algebraic study of quantifiers, where S5 modal logic is
  characterized by the class of all closure algebras in which each closed element is also open.
  In fact, the \emph{complex algebra} (over the set of possible worlds) of this S5 multi-modal logic FOL$_{\K}$ uses several S5 algebraic modal operators to provide a Boolean model features
  of FOL as indicated by Davis \cite{Davi54} in his doctoral thesis supervised by Garret Birkhoff.
  The Definition \ref{def:FOL-KripSem}, as supposed  by me, is the first attempt to give a relational Kripke
  semantics to the FOL, analogous to such an algebraic approach.\\
 As we can see from the definition of Kripke models of the
 FOL, every function and predicate are \emph{rigid} in it, that is, they have the
 same extension in every possible world $w \in \mathbb{W}$.
 \begin{theo} \label{Th:FOL-KripSem}
 The modal Kripke semantics in Definition \ref{def:FOL-KripSem} is an
 adequate semantics for the FOL.
 For each Tarski's interpretation $I_T$ there is an unique Kripke's
 interpretation $\M$ (exactly a Kripke model of FOL), and vice versa, such that for any $p_i^k \in
 P$, $f_i^k \in F$, $(d_1,...,d_k) \in \D^k$ and any intrinsic world (assignment) $g:Var \rightarrow \D$ it holds that:\\
 $I_K(g,p_i^k)(d_1,...,d_k) = t~~$ iff $~~(d_1,...,d_k) \in I_T(p_i^k)~~$ and\\
 $I_K(g,f_i^k)(d_1,...,d_k) = u~~$ iff $~~ u =
 I_T(f_i^k)(d_1,...,d_k)$.\\
 We define $\mathfrak{I}_K(\Gamma) = \{I_K$ defined above from $I_T~|~I_T \in \mathfrak{I}_T(\Gamma)\}$
  with the bijection $~\flat:\mathfrak{I}_T(\Gamma) \simeq
 \mathfrak{I}_K(\Gamma)$, so that for any Tarski's interpretation we
 have its equivalent Kripke's interpretation $I_K = \flat(I_T)$, where $\Gamma$ is a set of assumptions in this FOL. Moreover, the following commutative diagram of reductions
 is valid
 
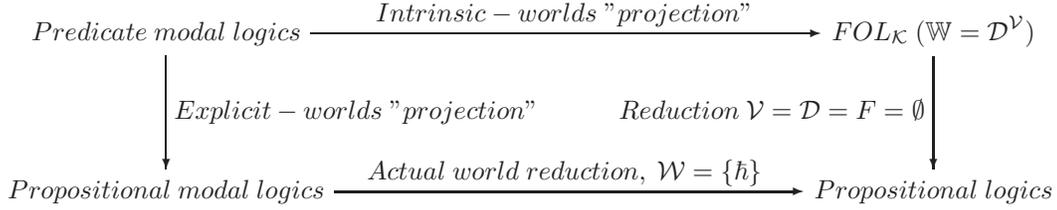
\begin{diagram}
   Predicate~ modal ~logics &    \rTo^{Intrinsic-worlds ~"projection"} &  FOL_{\K} ~(\mathbb{W} = \D^{\V})\\
  \dTo_{Explicit-worlds ~"projection"} &  &  \dTo^{Reduction ~\V = \D = F = \emptyset} \\
Propositional ~ modal~logics  & \rTo^{Actual~world~reduction, ~\W =
\{\hbar\}} & Propositional ~logics
\end{diagram}
 \end{theo}
 \textbf{Proof:} Let $(d_1,...,d_k) \in I_T(p_i^k)$, then from
 definition in this theorem $I_K(g,p_i^k)(d_1,...,d_k) = t$, and
 from Definition \ref{def:FOL-KripSem} we obtain that
 $|p_i^k(d_1,...,d_k)| = \mathbb{W}$, that is, the ground atom
 $p_i^k(d_1,...,d_k)$ is true also in correspondent Kripke's modal
 semantics. Viceversa, if $(d_1,...,d_k) \notin I_T(p_i^k)$, then $|p_i^k(d_1,...,d_k)| =
 \emptyset$ is a empty set, that is, the ground atom $p_i^k(d_1,...,d_k)$ is false also in correspondent Kripke's modal
 semantics. \\
 Let us suppose that
 for any  formula $\phi/g$ with $n$ logic connectives, true w.r.t. Tarski's interpretation $I_T$,  it holds
 that $|\phi/g| = \mathbb{W}$, that is, it is true in the correspondent Kripke's interpretation $I_K = \flat(I_T)$. Let us show that it holds for
 any formula $\psi$ with $n+1$ logic connectives,  true w.r.t. Tarski's interpretation $I_T$; there are the
 following three cases:\\
 1. $\psi = \phi_1 \wedge \phi_2$. Then $|\psi| = |\phi_1| \bigcap |\phi_2| = \mathbb{W}$,
 from the fact that both formulae $\phi_1, \phi_2$ must be true in Tarski's interpretation $I_T$ and that have less
 than or equal to $n$ logic connectives, and, consequently (by inductive assumption), $|\phi_i|  = \mathbb{W}$ for $i = 1,2$.
 That is, $\psi$ is true in the correspondent Kripke's interpretation $I_K = \flat(I_T)$.\\
  2. $\psi = \neg \phi$. Then $|\psi| = \mathbb{W} \backslash |\phi| = \mathbb{W}$, from the fact that the formula $\phi$ must be false in Tarski's interpretation $I_T$
  and that has less than or equal to $n$ logic connectives, and, consequently (by inductive assumption), $|\phi|  = \emptyset$.
  That is, $\psi$ is true in the correspondent Kripke's interpretation $I_K = \flat(I_T)$.\\
3. $\psi = (\exists x)\phi(x)$ where $\phi(x)$ denotes the formula
$\phi$ with the unique free variable $x$. From the fact that
$(\exists x)\phi(x)$ is true in Tarski's interpretation we have that
there is a value $u \in \D$ such that a sentence $\phi[x/u]$, that
is a formula $\phi$ where the variable $x$ is substituted by the
value $u \in \D$, is true
sentence. Then we obtain:\\
$|(\exists x)\phi(x)| = \{w~|~$exists $w'$ such that $(w,w') \in
\R_x$ and $~{\M} \models_{w'} \phi(x) \} = \mathbb{W}$,\\
because for any $w \in \D^{\V}$ there is $w'\in \D^{\V}$   such that
$w'(x) = u$ and for all $ y \in (\V \backslash \{x\})~ w'(y) =
w(y)$, and consequently $(w,w') \in \R_x$. It holds that  $~{\M}
\models_{w'} \phi(x)$, i.e., $\phi(x)$ is satisfied for the
assignment $w'$, because $w'(x) = u$ and $\phi(w'(x))$ is equivalent
to $\phi[x/u]$ which is true sentence with (from
inductive hypothesis)  $|\phi[x/u]| = \mathbb{W}$. \\
Consequently, any sentence which is true in Tarski's interpretation
$I_T$ is also true in the Kripke's interpretation $I_K =
\flat(I_T)$. Vice versa, for any sentence true in Kripke's
interpretation $I_K \in\mathfrak{I}_K(\Gamma)$ it can be analogously
shown that it is also true in the Tarski's interpretation $I_T =
\flat^{-1}(I_K)$, where $\flat^{-1}$ is inverse of the bijection
$\flat$. Thus, the Kripke's semantics given in Definition
\ref{def:FOL-KripSem} is an adequate semantics for the FOL.\\
Consequently, both "projections" in diagram above, where FOL$_{\K}$
denotes this adequate Kripke's version of the FOL (i.e., FOL where
the quantifiers $(\exists x)$ are
interpreted as \emph{modal} existential operators), are valid.\\
Let us show that also other two reductions into Propositional logics
are valid and render commutative the diagram  above:\\
1. Actual world reduction, when $\W = \{\hbar\}$: then, for this
unique explicit actual world $\hbar$ we have that we can have only
one (non empty) accessibility relation $\R_i = \{(\hbar,\hbar)\}$,
so that unique possible existential modal operator $\diamondsuit_i$,
of this modal propositional logic obtained by this reduction, is an
\emph{identity} operation: that is, obtained reduction is a
propositional logic without modal operators, i.e., it is a pure
propositional logic. In fact, we have that for any propositional
formula $\phi$ and the unique  explicit world $\hbar \in \W$,\\
$~{\M} \models_{\hbar} \diamondsuit_i\phi~$ iff $~$ exists $w' \in
\W$ such that $(\hbar, w') \in \R_i$ and $~{\M} \models_{w'}
\phi~$\\
iff $~{\M} \models_{\hbar} \phi$.
The Kripke's mapping $I_K:P \rightarrow \textbf{2}^{\W}$, for the
singleton set $\W = \{\hbar\}$ and the bijection
$\textbf{2}^{\{\hbar\}} \simeq \textbf{2}$, becomes the
propositional interpretation $I'_K:P \rightarrow \textbf{2}$.
Thus, we obtained a pure propositional logic in the actual world.\\
2. Reduction $~\V = \D = F = \emptyset$ from FOL$_{\K}$: from the
fact that $\V$ is the empty set of variables, we have that in $P$
all symbols become nullary symbols, that is propositional symbols,
so that the obtained logic is without modal operators (that is
without existential FOL quantifier $\exists$). Consequently, the
obtained logic is a propositional logic with  a unique generalized
world equal to
the empty function $*:\emptyset \rightarrow \D$, from the fact that
 $ \mathbb{W} =  \D^{\emptyset} = \{*\}$.
The Kripke's mapping $~I_K:\W\times P \rightarrow {\bigcup}_{n \in
\N} \textbf{2}^{\D^n}$ for FOL$_{\K}$ in this reduction becomes the
mapping $~I_K:\{*\}\times P \rightarrow \textbf{2}^{\D^0}$ where
$\D^0$ is the singleton set $\{<>\}$, so that, from bijections
$\{*\}\times P \simeq P$ and $\textbf{2}^{\{<>\}} \simeq
\textbf{2}$, this mapping becomes the propositional interpretation
$I'_K:P \rightarrow \textbf{2}$.
 We can consider the unique generalized world $*$ equivalent to the unique actual world, so that we obtain
exactly the same propositional logics in the actual world, as in the
case above.
\\$\square$\\
There is a surprising result from this theorem and its commutative
diagram: we obtained that a FOL (more precise its modal
interpretation of quantifiers in FOL$_{\K}$) is a particular
\emph{reduction} from the predicate modal logics. But it is well
known that the propositional modal logics can be, based on modal
correspondence theory \cite{Bent84,WaPe88}, embedded into the FOL by
transforming each propositional letter $p_i$ into an unary predicate
$p_i^1(x)$ (where $x$ is a new variable with domain of values equal
to the set of possible worlds of the original propositional modal
logic), and by introducing a binary predicate $\R(x,y)$ for the
accessibility relation of the Kripke semantics for propositional
modal operators.\\ For example, the (T) axiom $\Box p_i \Rightarrow
p_i$ of the propositional modal logic with universal modal operator
$\Box$ and with associated binary accessibility relation $\R$ over
the set of possible worlds, will be translated in the  FOL formula
$(\forall x)((\forall y)(\R(x,y) \Rightarrow p_i^1(y)) \Rightarrow
p_i^1(x))$. Analogously, the fact that a propositional letter $p_i$
is satisfied in the possible world $w$ in a given Kripke
interpretation $\M$, denoted by $~{\M} \models_{w} p_i~$, is
translated in the true
ground FOL atom $p_i^1(w)$.\\
Let us show that the similar embedding of the modal predicate logic
FOL$_{\K}$ into the FOL \emph{is not} possible. It is possible for
translation of satisfaction of the atoms of FOL$_{\K}$ in a possible
world $g \in \D^{\V}$, i.e. for $~{\M} \models_{g}
p^k_i(x_1,...,x_k)$, into a ground FOL atom
$p^k_i(g(x_1),...,g(x_k))$. But, for example, the translation of the
satisfaction of the modal formula of FOL$_{\K}$, i.e.  for $~{\M}
\models_{g} \diamondsuit_m p^k_i(x_1,...,x_m,...,x_k)$ where
$\diamondsuit_m$ is a modal interpretation of the syntax expression
$(\exists x_m)$, will be the formula $(\exists g')(\R_m(g,g') \wedge
p^k_i(g'(x_1),...,g'(x_m),...,g'(x_k))$. But it is a
\emph{second-order} formula, because $(\exists g')$ is a
quantification over functions (or equivalently over predicates that
represent the graphs of the assignment functions in $\D^{\V}$).\\
Thus, while the FOL can be equivalently translated by the modal
predicate logic FOL$_{\K}$, this modal predicate logic cannot be
equivalently translated in the FOL. That is, modal predicate logics
are more expressive than the FOL, and this is a surprising result,
at least for me. As we will see also the modal logic FOL$_{\K}$ is
an extensional logic as is FOL (they are equivalent logics), so that
also without enriching logics with the intensionality, the modal
predicate logics are still more expressive then the FOL.
\section{Modal logics and intensionality
\label{section:MODint}}
FOL$_{\K}$ represents the modal interpretation, with Kripke
relational semantics based on the set of possible worlds
$\mathbb{W}$, of the First-order logic. Consequently, based on
considerations in Section \ref{section:Intensionality} which
demonstrate that each modal logic with a set of possible worlds can
be considered as an intensional logic, we are invited to conclude
that also FOL is intrinsically an intensional logic. That is, by
introducing the particular structure of a domain $\D$ based on PRPs
we are able to define the intensional interpretation $I$ of the FOL
and the set of extensionalization functions $h = is(w) \in \E$ for
any possible world $w \in \mathbb{W} = \D^{\V}$ in the FOL$_{\K}$
Kripke semantics of the FOL.  But in the case of modal FOL$_{\K}$ we
will see that the intensions of logic formulae are equivalent to
their extensions, that is, the original FOL also with Kripke
semantics (equivalent to Tarski's semantics) is not able to support
the intensionality differently from the extensionality.
\begin{propo} \label{prop:FOL-KripSem} The intension (sense) of any virtual predicate $\phi(x_1,...,x_k)$
in the  FOL$_{\K}$, with a set of only intrinsic possible worlds
$\mathbb{W} = \D^{\V}$, is equivalent to its extension in a given
Tarski's interpretation of the FOL. That is, it is impossible to
support the intensions in the standard FOL with Tarskian semantics.
\end{propo}
\textbf{Proof:} We have to show that the intension
$I_n(\phi(x_1,...,x_k)):\mathbb{W} \rightarrow \mathfrak{R}$ of any
FOL formula $\phi(x_1,...,x_k)$ in a given Tarski's interpretation
$I_T^*:\L\rightarrow \mathfrak{R}$ is a constant function from the
set of possible worlds $\mathbb{W} = \D^{\V}$ in FOL$_{\K}$, such
that for all  $w \in \mathbb{W}$ we have that
$I_n(\phi(x_1,...,x_k))(w) = R$, where $R =
I_T^*(\phi(x_1,...,x_k))$ is the extension of this formula in this
Tarski's interpretation. We can show it by the
structural recursion:\\
 1. Case when $\phi(x_1,...,x_k)$ is a predicate letter $p_i^k \in
 P$. Then,  \\$I_n(p_i^k(x_1,...,x_k))(w) = \|p_i^k(x_1,...,x_k)\|_{\M, w}\\ =
 \{ (d_1,...,d_k) \in \D^k ~| ~  {\M}
\models_{w}~ p_i^k(d_1,...,d_k) \} \\
= \{ (d_1,...,d_k) ~|~ I_K(w,p_i^k)(d_1,...,d_k) = t\} \\
=  \{ (d_1,...,d_k) ~|~ (d_1,...,d_k) \in I_T(p_i^k) \} = I_T(p_i^k)$.\\
2. Case when $\phi(x_1,...,x_k)$ is a virtual predicate. From
Theorem \ref{Th:FOL-KripSem} it holds that if for a given assignment
$g \in \mathbb{W}$ a ground formula $\phi(x_1,...,x_k)/g$ is true in
a given Tarski's interpretation $I_T$ (i.e., when
$I_T^*(\phi(x_1,...,x_k)/g) = t$), then $|\phi(x_1,...,x_k)/g| =
\mathbb{W}$ (it is true in the corespondent Kripke's
interpretation), that is, $~{\M} \models_{w} \phi(x_1,...,x_k)/g$
for \emph{every} possible world $w \in \mathbb{W}$. Thus, we have
that the intension of this virtual predicate is,
$~I_n(\phi(x_1,...,x_k))(w) = \|\phi(x_1,...,x_k)\|_{\M, w}\\ =
 \{ (g(x_1),...,g(x_k)) \in \D^k ~| ~g \in \mathbb{W}$ and  ${\M}
\models_{w}~ \phi(x_1,...,x_k)/g \}  \\
= \{ (g(x_1),...,g(x_k))  ~| ~g \in \mathbb{W}$ and
$I_T^*(\phi(x_1,...,x_k)/g) = t \}\\
= \{ (g(x_1),...,g(x_k))  ~| ~g \in \mathbb{W}$ and
$(g(x_1),...,g(x_k)) \in I_T^*(\phi(x_1,...,x_k)) \} \\=
I_T^*(\phi(x_1,...,x_k))$. Thus, the function $I_n$ is invariant
w.r.t. the possible worlds $w \in \mathbb{W}$, and returns with the
extension, of a considered (virtual) predicate, determined by a
given Tarski's interpretation.
\\$\square$\\
What does it mean? First of all it means that not every modal logic
with a given set of possible worlds $\mathbb{W}$ is an intensional
logic, and that the quality of the intensionality which can be
expressed by a given modal logic depends on the set of possible
worlds $\mathbb{W}$ and their capacity to model the possible
extensions of logic formulae. For example, if $\mathbb{W}$ is a
finite set with very small cardinality, it often wold not be able to
express the all possible extensions for logic formulae, and,
consequently, its intensional capability will be very limited. But
also if $\mathbb{W}$ is infinite, as in the case above when $\D$ is
an infinite domain,  we demonstrated that they are not able to
express the intensionality. Consequently, in order to be able to
express the full intensionality in a given modal logic, it is very
important to chose the new \emph{appropriate set} of possible
worlds, independently from the original set of possible worlds of
the particular given modal logic.\\
In fact, from this point of view, the left arrow in the diagram in
Theorem \ref{Th:FOL-KripSem} represents the logics with (partial)
intensionalities, while the right arrow of the same diagram
represents two extremal reductions of the intensionality, by
identifying it with the pure extensionality (the propositional logic
can be seen as a modal logic with the unique actual possible worlds,
so that the intensionality corresponds to the extensionality, as in
the case of
the FOL$_{\K}$).\\
 \textbf{Remark:} The natural
choice for the set of explicit possible worlds for the fully
intensional logic is the set of interpretations of its original
logic (modal or not, determined by its set of axioms, inference
relations, and a predefined set $\Gamma$, possibly empty, for which
these interpretations are models), because such a set of
interpretations is able to express the all logically possible
extensions of the formulae of the original (not fully intensional)
logic. In what follows we will do this intensional upgrade for the
standard (not modal) FOL$(\Gamma)$, but generally it can be done to
every kind of logics, thus to any kind of modal logics, consequently
also to the modal logic FOL$_{\K}(\Gamma)$: in that case we obtain
the two-levels modal logic (as in \cite{Majk06S,Majk08in}). At the
lower-level we will have original modal logics with their original
set of possible worlds (the set $\D^{\V}$ in the case of
FOL$_{\K}(\Gamma)$), while at the new upper-level each new explicit
possible world would correspond to the particular Kripke's
interpretations of the original modal logics. The obtained
upper-level intensional logic has a kind of \emph{rigid} semantics,
where the domains  and the extensions of built-in
predicates/propositions  of the "lover-level" modal logics are
identical in every upper-level possible world.
 \\$\square$\\
In intensional logics a k-ary functional symbol $f^k_i \in F$ is
 considered as the new $k+1$-ary "functional" predicate symbol $f^{k+1}_i \in P$ whose extension  is
 the graph of this function, such that cannot exists two tuples $(d_1,...,d_k,u_1), (d_1,...,d_k,u_2)$
  in its extension with $u_1\neq u_2$ (by introducing new axiom $(\exists_1 x_{k+1})f^{k+1}_i(x_1,...,x_{k+1})$). Thus, in what follows we will have only
 the set of predicate symbols.\\
This two-level intensional modal logic with the orthogonality of old
possible worlds of the original modal logic $\mathbb{W} = \W \times
\D^{\V}$ and the new set of explicit possible worlds
$\mathfrak{I}_K$ (the set of all Kripke interpretations $I_K \in
\mathfrak{I}_K(\Gamma)$,  of the original (non intensional) modal
logic, in which all assumptions in $\Gamma$ (possibly empty set) are
true), means that the obtained intensional modal logic has the set
of explicit possible worlds equal to the cartesian product of old
explicit worlds $\W$ and new added worlds in
$\mathfrak{I}_K(\Gamma)$, so that new generalized possible worlds
are equal to the set $\widehat{\mathbb{W}} = (\mathfrak{I}_K(\Gamma)
\times \W) \times \D^{\V}$. Consequently, the Kripke semantics of
fully intensional modal logic, obtained as an enrichment of the
original modal logic, can be given by the following definition:
\begin{definition} \label{def:IntKripSem} \textsc{Intensional enrichment of multi-modal logics}:\\
 Let $\M = (\mathbb{W}, \{
{\R}_i \}, \D, I_K)$ be a   Kripke's interpretation of an original
multi-modal logic with the set of (generalized) possible worlds
$\mathbb{W} = \W \times \D^{\V}$ and the set of existential modal
operators $\diamondsuit_i$ with accessibility relations $\R_i$,
given by
Definition \ref{def:NewKripSem}.\\
Then we denote by $\widehat{\M} = (\widehat{\mathbb{W}}, \{ {\R}_i
\}, \{ {\widehat{\R}}_j\}, \D, \widehat{I}_K)$ a Kripke's
interpretation of its intensional enrichment with the set of
possible worlds $\widehat{\mathbb{W}} = \mathfrak{I}_K(\Gamma)
\times \mathbb{W}$, the optional set of new \verb"intensional" modal
operators $\widehat{\diamondsuit}_j$ with the accessibility
relations $\widehat{\R}_j$  over the worlds in
$\mathfrak{I}_K(\Gamma)$, and new mapping
$~~\widehat{I}_K:(\mathfrak{I}_K(\Gamma) \times \W) \times P
\rightarrow {\bigcup}_{n \in \N} \textbf{2}^{\D^n}$, such that for
any explicit world $(I_K,w) \in \mathfrak{I}_K(\Gamma) \times \W$
and $p_i^k \in P$ we have that $~\widehat{I}_K(I_K,w,p_i^k) =_{def}
I_K(w,p_i^k):\D^k \rightarrow \textbf{2}$. The satisfaction relation
$\models_{I_k, w,g}$ for a given world  $(I_k, w,g) \in
\widehat{\mathbb{W}}$ is defined as follows:\\
1. $~~{\widehat{\M}} \models_{I_K,w,g}~p_i^k(x_1,...,x_k)~~~$ iff $~~~\widehat{I}_K(I_K,w,p_i^k)(g(x_1),...,g(x_k)) = t$.\\
2. $~~{\widehat{\M}} \models_{I_K,w,g}~ \varphi \wedge \phi~~~$ iff
$~~~{\widehat{\M}} \models_{I_k,w,g}~ \varphi~$ and $~{\widehat{\M}}
\models_{I_K,w,g}~ \phi~$, \\
3. $~~{\widehat{\M}} \models_{I_K,w,g}~ \neg \varphi ~~~$ iff $~~~$ not ${\widehat{\M}} \models_{I_K,w,g}~ \varphi~$, \\
4.  $~~{\widehat{\M}} \models_{I_K,w,g}~\lozenge_i \varphi~~~$ iff
$~~~$ exists $w'\in \W$ such that $(w,w')
\in {\R}_i $ and ${\widehat{\M}} \models_{I_K,w',g}~ \varphi$.\\
5.  $~~{\widehat{\M}} \models_{I_K,w,g}\widehat{\lozenge}_j
\varphi~~$ iff $~~$ exists $I_K'\in \mathfrak{I}_K$ such that
$(I_K,I'_K) \in {\widehat{\R}}_j $ and ${\widehat{\M}}
\models_{I'_K,w,g}~ \varphi$.
 \end{definition}
 Notice that this intensional enrichment is \emph{maximal} one: in fact we
 have taken \emph{all}  Kripke's interpretations of the original modal
 logics for the possible worlds of this new intensional logic. We can
 obtain partial intensional enrichments if we take only a strict
 subset of $S \subset \mathfrak{I}_K(\Gamma)$ in order to define generalized
 possible worlds $\widehat{\mathbb{W}} = S \times
\mathbb{W}$. In that case we would introduce the non monotonic property for obtained intensional logic.\\
 \textbf{Example 1:} Let us consider the intensional
enrichment of the multi-modal logic FOL$_{\K}(\Gamma)$ given by
Definition \ref{def:FOL-KripSem} with the Kripke's interpretation
$\M = (\mathbb{W}, \{ {\R}_x~ | ~x \in \V \}, \D, I_K)$ of the
FOL$(\Gamma)$, with
 a set of (generalized) possible worlds   $ \mathbb{W} = \D^{\V}$
 and  the accessibility relation
 ${\R}_x = \{(w_1,w_2) \in \mathbb{W} \times \mathbb{W}~ | ~x\in \V$ and for all $y \in \V \backslash \{x\}(w_1(y) = w_2(y))\}$
 for existential modal operator $(\exists x)$ for each variable $x \in
\V$.\\
Then  $\widehat{\M} = (\widehat{\mathbb{W}}, \{ {\R}_x \}, \{
{\widehat{\R}}_j\}, \D, \widehat{I}_K)$ is a Kripke's interpretation
of its intensional enrichment with the set of generalized possible
worlds $\widehat{\mathbb{W}} = \mathfrak{I}_K(\Gamma) \times
\mathbb{W} = \mathfrak{I}_K \times \D^{\V}$ (here the set of
explicit worlds is $\mathfrak{I}_K(\Gamma)$), the optional set of
new modal operators $\widehat{\diamondsuit}_j$ with the
accessibility relations $\widehat{\R}_j$  over the worlds in
$\mathfrak{I}_K(\Gamma)$, and new mapping
$~~\widehat{I}_K:(\mathfrak{I}_K(\Gamma) \times \D^{\V}) \times P
\rightarrow {\bigcup}_{n \in \N} \textbf{2}^{\D^n}$, such that for
any explicit world $(I_K,g) \in \mathfrak{I}_K(\Gamma) \times
\D^{\V}$ and $p_i^k \in P$ we have that $~\widehat{I}_K(I_K,g,p_i^k)
=_{def}
I_K(g,p_i^k):\D^k \rightarrow \textbf{2}$, with\\
$~~{\widehat{\M}} \models_{I_K,g}~\lozenge_x \varphi~~~$ iff $~~~$
exists $g'\in \D^{\V}$ such that $(g,g') \in {\R}_x $ and
${\widehat{\M}} \models_{I_K,g'}~ \varphi$.\\
Then, from Definition \ref{def:intensemant} for the intensional
semantics, the mapping $I_n:\L_{op} \rightarrow
\mathfrak{R}^{\mathfrak{I}_K(\Gamma)}$, where $\L_{op}$ is the
subset of formulae with free variables (virtual predicates), such
that for any virtual predicate $\phi(x_1,...,x_k) \in \L_{op}$ the
mapping $I_n(\phi(x_1,...,x_k)):\mathfrak{I}_K(\Gamma) \rightarrow
\mathfrak{R}$ is the Montague's meaning (\emph{intension}) of this
virtual predicate, i.e. mapping which returns with the extension of
this predicate in every explicit possible world (i.e., Kripke's
interpretation of
FOL$_{\K}(\Gamma)$) $I_K \in \mathfrak{I}_K(\Gamma)$. That is, we have that\\
$I_n(\phi(x_1,...,x_k))(I_K) =_{def}  \|\phi(x_1,...,x_k)\|_{\M, I_K}\\
=
 \{ (g(x_1),...,g(x_k)) \in \D^k ~| ~g \in \D^{\V}$ and  ${\widehat{\M}}
\models_{I_K,g}~ \phi(x_1,...,x_k) \}$.\\
In what follows, the \emph{minimal} (i.e. without new intensional
modal operators $\widehat{\diamondsuit}_i$) intensional enrichment
of the multi-modal logic FOL$_{\K}$ we will denote by
FOL$_{\K_{\I}}$.
\\$\square$\\
This two-level intensional modal logic, described above, has the
following correspondence property between the Kripke's
interpretation $\M$ of the original modal logic and the Kripke's
interpretation of $\widehat{\M}$ its intensional enrichment:
\begin{propo} \label{prop:IntKripSem}
For any logic formulae $\phi$ of the original multi-modal logic,
with the set of (generalized) possible worlds $\mathbb{W} = \W
\times \D^{\V}$ and the set of existential modal operators
$\diamondsuit_i$ with accessibility relations $\R_i$ given by
Definition \ref{def:NewKripSem}, the following property is valid:
$~~{\widehat{\M}} \models_{I_K,w,g}~ \phi~~~$ iff $~~~\M
\models_{w,g}~ \phi$,\\
where $\M = (\mathbb{W}, \{ {\R}_i \}, \D, I_K)$ is a   Kripke's
interpretation of the original multi-modal logic. Consequently,
$\phi$ is true in the intensionally enriched multi-modal logic iff
it is \verb"valid" in the original multi-modal logic.
\end{propo}
\textbf{Proof:} Let us demonstrate it by structural induction on the
length of logic formulae. For any atom  $\phi = p_i^k(x_1,...,x_k)$
we have from Definition \ref{def:IntKripSem} that $~~{\widehat{\M}}
\models_{I_K,w,g}~p_i^k(x_1,...,x_k)~~~$ iff
$~~~\widehat{I}_K(I_K,w,p_i^k)(g(x_1),...,g(x_k)) =
I_K(w,p_i^k)(g(x_1),...,g(x_k)) = t ~~~$ iff \\$~~~\M \models_{w,g}~
p_i^k(x_1,...,x_k)$. Let us suppose that such a property holds for
every formula $\phi$ with less than $n$ logic connectives of the
original multi-modal logic (thus without new intensional connectives
$\widehat{\diamondsuit}_i$), and let us show that it holds also for
any formula with $n$ logic connectives. There are the following
cases:\\
1. The case when $\phi = \neg \psi$ where $\psi$ has $n-1$ logic
connectives. Then $~~{\widehat{\M}} \models_{I_K,w,g}~ \phi~~~$ iff
$~~~{\widehat{\M}} \models_{I_K,w,g}~ \neg \psi ~~~$ iff  $~~~$ not
${\widehat{\M}} \models_{I_K,w,g}~ \psi ~~~$ iff (by inductive
hypothesis) $~~~$ not $\M \models_{w,g}~ \psi ~~~$ iff $~~~\M
\models_{w,g}~ \neg \psi ~~~$ iff $~~~\M \models_{w,g}~ \phi$.\\
 2. The case when $\phi =  \psi_1 \wedge \psi_2$, where both $\psi_1,
\psi_2$ have less than $n$ logic connectives, is analogous to the
case 1.\\
3. The case when $\phi = \diamondsuit_i \psi$ where $\psi$ has $n-1$
logic connectives. Then $~~{\widehat{\M}} \models_{I_K,w,g}~
\phi~~~$ iff $~~~{\widehat{\M}} \models_{I_K,w,g}~ \diamondsuit_i
\psi ~~~$ iff  $~~~$ exists $w', (w,w') \in \R_i$ and
${\widehat{\M}} \models_{I_K,w,g}~ \psi ~~~$ iff (by inductive
hypothesis)  $~~~$ exists $w', (w,w') \in \R_i$ and $\M
\models_{w,g}~ \psi ~~~$ iff $~~~\M \models_{w,g}~ \diamondsuit_i
\psi ~~~$ iff $~~\M \models_{w,g}~ \phi$.
\\$\square$

\section{First-order logic and intensionality
\label{section:FOLint}}
 Thus,
in order to be able to manage the intensions of logic formulae, our
modal Kripke semantics for the FOL$(\Gamma)$ has to be enriched also
by a set of explicit possible worlds where each predicate can have
\emph{different} extensions.
  Such an \emph{intensional semantics} of FOL$(\Gamma)$ is strictly more
expressive than a single Tarskian semantics of FOL$(\Gamma)$: as we
have seen in Section \ref{section:Intensionality}, the intensional
semantics of FOL$(\Gamma)$ with a set of logic formulae $\L$, given
by the composed mapping $~ \L ~\longrightarrow_I~ \D
~\Longrightarrow_{w \in \W}~ \mathfrak{R}~$ in Definition
\ref{def:intensemant}, is equivalent to \emph{all} Tarski's
interpretations of FOL$(\Gamma)$, where for each Tarski's
interpretation $I_T$ (and its unique extension $I_T^*$ to all
formulae in $\L$) we obtain a single extensionalization function $h
\in \E$ such that $I_T^* = h \circ I$.\\
In fact, with a given standard Tarski's interpretation $I_T$ of
FOL$(\Gamma)$ we are not able to express the \emph{intensional
equality} of two open formulae with the same tuple of free
variables, $\phi(x_1,..x_n), \psi(x_1,..x_n)$, defined by $(\forall
h)
(h(I(\phi(x_1,..x_n))) = h(I(\psi(x_1,..x_n))))$.\\
 From the fact that any predicate in FOL$(\Gamma)$ can have different extensions only
 for the set of different Tarski's interpretations of FOL$(\Gamma)$,
 the natural choice for explicit worlds is the
 set of all Tarski's interpretations of FOL$(\Gamma)$ denoted by $\mathfrak{I}_T(\Gamma)$, i.e. $\W = \mathfrak{I}_T(\Gamma)$, so that the set of
 generalized possible worlds in obtained intensional semantics is
 equal to $\mathbb{W} = \mathfrak{I}_T(\Gamma) \times \D^{\V}$.\\
  Two virtual predicates with the same tuple of free variables, $\phi(x_1,..x_n)$ and
 $\psi(x_1,..x_n)$, are \emph{intensionally equal} iff the
 formula  $\phi(x_1,..x_n) \equiv
 \psi(x_1,..x_n)$ is true in this FOL$_{\I}(\Gamma)$ (thus, satisfied in every explicit possible world, that is, in every Tarski's
 interpretation of FOL$(\Gamma)$).\\
  Let us define this \emph{minimal intensional} first-order logic FOL$_{\I}(\Gamma)$ which has the same syntax as
 standard FOL (thus without other (modal) logic connectives), but  enriched with the set
  of possible worlds $\mathbb{W} = \mathfrak{I}_T(\Gamma) \times \D^{\V}$.
 \begin{definition} \label{def:IntensionalFOL} \textsc{Minimal Intensional First-order Logic (FOL$_{\I}(\Gamma)$):}\\
 We denote by $\M_{FOL_{\I}(\Gamma)} = (\mathbb{W},
 \D, I_K)$ the Kripke's interpretation of the Intensional logic
 FOL$_{\I}(\Gamma)$
 with
 a set of (generalized) possible worlds $\mathbb{W}$, a set of explicit possible
 worlds equal to the set of Tarski's interpretation of FOL$(\Gamma)$,
 $\W = \pi_1(\mathbb{W})= \mathfrak{I}_T(\Gamma)$ and $ \pi_2(\mathbb{W}) = \D^{\V}$,
  non empty domain $\D$, and  the mapping  $~~I_K:\W\times P \rightarrow {\bigcup}_{n \in \N}
\textbf{2}^{\D^n}$.\\
We extend the satisfaction relation  $\models_{w,g}$ of Kripke
semantics to the first-order quantification $\exists$ by:
$~~\M_{FOL_{\I}(\Gamma)} \models_{w,g}~
(\exists x) \phi ~~$ iff \\
1. $~~\M_{FOL_{\I}(\Gamma)} \models_{w,g}~ \phi$, if $x$ is not a
free
variable in $\phi$;\\
2. $~~$exists $u \in \D$ such that $~~\M_{FOL_{\I}(\Gamma)}
\models_{w,g}~ \phi[x/u]$, if $x$ is  a free variable in $\phi$ and
$\phi[x/u]$ the formula obtained by substitution of $x$ by the value
$u$ in $\phi$.\\
 Such an interpretation is the Kripke \verb"model" of
Intensional FOL if for any explicit world (Tarski's interpretation)
$w = I_T \in \W$, $p_i^k \in P$, and a tuple
$(d_1,...,d_k) \in \D^k$, we have that:\\
$~I_K(w,p_i^k)(d_1,...,d_k) = t~$ iff $~(d_1,...,d_k) \in w(p^k_i)$.
 \end{definition}
 Notice that the intensional semantics above is given for the
ordinary syntax of the First-order logic with the existential
quantifier $\exists$, without modal operators, thus with the empty
set of accessibility binary relations over the set of explicit
possible worlds $\W =
\pi_1(\mathbb{W})= \mathfrak{I}_T(\Gamma)$: this is the reason to denominate it by "minimal". \\
 Let us show that this unique intensional Kripke model $\M_{FOL_{\I}(\Gamma)} = (\mathbb{W},
 \D, I_K)$  models the Tarskian \emph{logical consequence} of the
 First-order logic with a set of assumption in $\Gamma$, so that the added intensionality preserves the Tarskian semantics of the FOL.
%
\begin{propo} \label{prop:IntensionalFOL}
Let $\M_{FOL_{\I}(\Gamma)} = (\mathbb{W},
 \D, I_K)$ be the  unique \verb"intensional" Kripke model  of the
 First-order logic with a set of assumptions in $\Gamma$, as defined
 in Definition \ref{def:IntensionalFOL}.\\
Then, a formula $\phi$  is \emph{a logical consequence} of $\Gamma$
in the Tarskian semantics for the FOL,
  that is, $~~\Gamma\Vdash \phi,~~$ iff $~~\phi$ is true in this Kripke intensional model
  $\M_{FOL_{\I}(\Gamma)}$.\\
 Let $I_n:\L_{op} \rightarrow
\mathfrak{R}^{\W}$ be the mapping given in Definition
\ref{def:intensemant}. Then, for any (virtual) predicate
$\phi(x_1,...,x_k) $, the mapping $I_n(\phi(x_1,...,x_k)):\W
\rightarrow \mathfrak{R}$  represents the Montague's meaning
(intension) of this logic formula, such that:\\ for any $w \in \W =
\pi_1(\mathbb{W})$, $~~I_n(\phi(x_1,...,x_k))(w) =
w^*(\phi(x_1,...,x_k))$.
 \end{propo}
\textbf{Proof:} Let us show that for any first-order formula $\phi$
it holds that, $~~\M_{FOL_{\I}(\Gamma)} \models_{w,g}~ \phi ~~$ iff
$~~w^*(\phi/g) = t$, where $w^*$ is the unique extension of Tarski's
interpretation $w = I_T \in \W = \mathfrak{I}_T(\Gamma)$ to all
formulae.\\Let us demonstrate it by the structural induction on the
length of logic formulae. For any atom $\phi = p_i^k(x_1,...,x_k)$
we have from Definition \ref{def:IntKripSem} that
$~~\M_{FOL_{\I}(\Gamma)} \models_{I_T,g}~p_i^k(x_1,\\...,x_k)~~~$
iff $~I_K(I_T,p_i^k)(g(x_1),...,g(x_k)) = t~~~$ iff
$~~~(g(x_1),...,g(x_k)) \in I_T(p^k_i)~~~$ iff
$~~~I_T(p_i^k(x_1,...,x_k)/g) = t$. Let us suppose that such a
property holds for every formula $\phi$ with less than $n$ logic
connectives of the FOL, and let us show that it holds also for any
formula with $n$ logic connectives. There are the following
cases:\\
1. The case when $\phi = \neg \psi$ where $\psi$ has $n-1$ logic
connectives. Then $~~\M_{FOL_{\I}(\Gamma)} \models_{I_T,g}~ \phi~~~$
iff $~~~\M_{FOL_{\I}(\Gamma)} \models_{I_T,g}~ \neg \psi ~~~$ iff
$~~~$ not $\M_{FOL_{\I}(\Gamma)} \models_{I_T,g}~ \psi ~~~$ iff (by
inductive
hypothesis) $~~~$ not $I_T^*(\psi/g) = t~~~$ iff  $~~~I_T^*(\neg\psi/g) = t~~~$ iff  $~~~I_T^*(\phi/g) = t$.\\
 2. The case when $\phi =  \psi_1 \wedge \psi_2$, where both $\psi_1,
\psi_2$ have less than $n$ logic connectives, is analogous to the
case 1.\\
3. The case when $\phi = (\exists x) \psi$ where $\psi$ has $n-1$
logic connectives. It is enough to consider the case when $x$ is a
free variable in $\psi$. Then $~~\M_{FOL_{\I}(\Gamma)}
\models_{I_T,g}~ \phi~~~$ iff $~~~\M_{FOL_{\I}(\Gamma)}
\models_{I_T,g}~ (\exists x) \psi ~~~$ iff $~~~$ exists $u \in \D$
such that $~~\M_{FOL_{\I}(\Gamma)} \models_{I_T,g}~ \psi[x/u]~~~$
iff (by inductive hypothesis)  $~~~$ exists $u \in \D$
such that $~~I_T^*(\psi[x/u]/g) = t~~~$  iff $~~~I_T^*((\exists x)\psi/g) = t ~~~$ iff $~~I_T^*((\phi/g) = t$.\\
It is easy to verify that the intension of predicates in the
FOL$_{\I}(\Gamma)$ defined above can be expressed by the mapping
$I_n$ such that for any $p^k_i \in P$, $I_n(p_i^k(x_1,...,x_k)(w) =
w(p_i^k)$, and, more
general, for any virtual predicate $\phi(x_1,...,x_k)$,\\
 $I_n(\phi(x_1,...,x_k))(w) =
\|\phi(x_1,...,x_k)\|_{\M, w}\\ =
 \{ (g(x_1),...,g(x_k)) \in \D^k ~| ~g \in \D^{\V}$ and  ${\M}_{FOL_{\I}}
\models_{w,g}~ \phi(x_1,...,x_k) \}  \\
= \{ (g(x_1),...,g(x_k))  ~| ~g \in \D^{\V}$ and
$w^*(\phi(x_1,...,x_k)/g) = t \}\\
= \{ (g(x_1),...,g(x_k))  ~| ~g \in \D^{\V}$ and
$(g(x_1),...,g(x_k)) \in w^*(\phi(x_1,...,x_k)) \} \\=
w^*(\phi(x_1,...,x_k))$,\\
where $w^*$ is the unique extension of Tarski's interpretation $w
\in \W = \mathfrak{I}_T(\Gamma)$ to all formulae. Consequently,
$I_n(\phi(x_1,...,x_k)):\W \rightarrow \mathfrak{R}$ is the
Montague's meaning (i.e., the intension) of the (virtual) predicate
$\phi(x_1,...,x_k)$.
\\$\square$\\
It is clear that in Kripke semantics of this intensional
 first-order logic, denoted by FOL$_{\I}(\Gamma)$, if the set of
 assumptions is empty ($\Gamma = \emptyset$), then
 a formula $\phi$ is true  in the intensional Kripke model $\M_{FOL_{\I}(\Gamma)}~~$ iff $~~$ it is \emph{valid}  in Tarskian semantics of the FOL,
  that is, iff $~\Vdash \phi$ in the FOL.\\
The main difference between Tarskian semantics and this intensional
semantics is that this unique intensional Kripke model
$\M_{FOL_{\I}(\Gamma)}$ encapsulates the set of all Tarski models of
the First-order logic with a (possibly empty) set of assumptions
$\Gamma$.
 \begin{coro} \label{coro:IntKripSem}
The intensionalities of two different minimal intensional
enrichments of the first-order syntax, given by intensional logics
FOL$_{\I}(\Gamma)$ and FOL$_{\K_{\I}}(\Gamma)$ (in Example 1), are
equivalent and correspond to Montague's intensionality.
\end{coro}
\textbf{Proof:} Let us denote by $I_n^{FOL_{\I}(\Gamma)},
I_n^{FOL_{\K_{\I}}(\Gamma)}:\L_{op} \rightarrow \mathfrak{R}^{\W}$
the intensional mappings  (from Definition \ref{def:intensemant} of
the intensional semantics) for these two intensional enrichments of
the FOL$(\Gamma)$. Notice that the set of  explicit possible worlds
$\W$ in FOL$_{\I}(\Gamma)$ is equal to $ \mathfrak{I}_T(\Gamma)$
while in FOL$_{\K_{\I}}(\Gamma)$ is equal to $
\mathfrak{I}_K(\Gamma)$, with the bijection (from Theorem
\ref{Th:FOL-KripSem}) $\flat:\mathfrak{I}_T(\Gamma) \simeq
\mathfrak{I}_K(\Gamma)$.
 We have to
show that for any formulae $\phi(x_1,...,x_n) \in \L_{op}$ its
extension, in a given explicit world $I_T \in \mathfrak{I}_T
(\Gamma)$ of the intensional logic FOL$_{\I}(\Gamma)$, is equal to
its extension in the correspondent explicit world $I_K = \flat(I_T)
\in \mathfrak{I}_K(\Gamma) $ of the intensional logic
FOL$_{\K_{\I}}(\Gamma)$.
In fact, we have that:\\
$I_n^{FOL_{\I}(\Gamma)}(\phi(x_1,...,x_k))(I_T) =
\|\phi(x_1,...,x_k)\|_{\M_{FOL_{\I}(\Gamma)}, I_T}\\ =
 \{ (g(x_1),...,g(x_k)) \in \D^k ~| ~g \in \D^{\V}$ and  $\M_{FOL_{\I}(\Gamma)}
\models_{I_T,g}~ \phi(x_1,...,x_k) \}  \\
= \{ (g(x_1),...,g(x_k))  ~| ~g \in \D^{\V}$ and
$I_T^*( \phi(x_1,...,x_k)/g) = t \}$ (from Prop. \ref{def:IntensionalFOL})  \\
$= \{ (g(x_1),...,g(x_k)) ~| ~g \in \D^{\V}$ and $\M \models_{g}~
\phi(x_1,...,x_k) \}$ (from Prop. \ref{prop:FOL-KripSem}
 and Theorem \ref{Th:FOL-KripSem} where $\M$ is the Kripke interpretation of FOL$_{\K}(\Gamma)$ with  $I_K = \flat(I_T)$)   \\
$= \{ (g(x_1),...,g(x_k))  ~| ~g \in \D^{\V}$ and
$\widehat{\M}_{FOL_{\K}(\Gamma)} \models_{I_K,g}~ \phi(x_1,...,x_k)
\}$
(from Prop. \ref{prop:IntKripSem})\\
$= \|\phi(x_1,...,x_k)\|_{\M_{FOL_{\K_{\I}}(\Gamma)}, I_K} =
I_n^{FOL_{\K_{\I}}(\Gamma)}(\phi(x_1,...,x_k))(I_K)$.
\\$\square$\\
That is, independently on how we interpret the quantifiers of the
FOL, as in standard FOL or as modal operators in FOL$_{\K}$, the
intensionality of the FOL is obtained only by one adequate
\emph{semantic enrichment}, without modifying its syntax.
Consequently, we have demonstrated that an intensional FOL does not
need the other logic operators as required by Bealer \cite{Beal82},
that is, we do not need intensional abstraction operator or another
modal operator. Because of that we denominated such an intensional
FOL as the \emph{minimal} intensional logic. Another intensional FOL
without
the intensional abstraction is given in the following example:\\
\textbf{Example 2:} In order to be able to recognize the intensional
equivalence
 between (virtual) predicates, that may be used in intensional mapping between P2P databases
 \cite{Majk04ph,Majk08in,Majk09e}, we need to extend this \emph{minimal} intensional FOL   also syntactically, by introducing the new
 modal existential  operator $\diamondsuit$, so that $\phi(x_1,..x_n)$ and
 $\psi(x_1,..x_n)$ are \emph{intensionally equivalent} iff the modal First-order
 formula  $\diamondsuit \phi(x_1,..x_n) \equiv
 \diamondsuit \psi(x_1,..x_n)$ is true in this modal FOL. The Kripke semantics for this extended modal first-order logic is a S5
 modal FOL with the accessibility relation ${\R} = \W \times  \W$.\\
  Two intensional \emph{equivalent} predicates does need to have
 equal extensions in each explicit possible world as is required by
 intensional \emph{equality} (equal meaning from Montague's point of
 view) when $\phi(x_1,..x_n) \equiv \psi(x_1,..x_n)$ is true,
 where
 '$\equiv$' is the standard logic equivalence
 connective.\\
 Notice that if they are intensionally equal, it does not mean that
 they are equal concepts, i.e that
 $I(\phi(x_1,..x_n)) = I(\phi(x_1,..x_n)) \in \D$, but only that
 they are necessarily equivalent. In fact, the two atoms  $p_1^1(x)$, "$x$ has
 been bought", and $p_2^1(x)$, "$x$ has
 been sold", are necessarily equivalent, that is, it holds that
 $p_1^1(x)\equiv p_2^1(x)$ but they are two different \emph{concepts}, that
 is $I(p_1^1(x)) \neq  I(p_2^1(x))$ (i.e., $(I(p_1^1(x)),I(p_2^1(x))) \notin h(Id) = R_=$). Such an distinction of equal
 concepts and of the intensional equality (i.e., the necessary equivalence)
 is not possible in the Montague's semantics, and explain why we
 adopted PRP theory and two-step intensional semantics in Definition \ref{def:intensemant} analogously
 to  Bealer's approach.\\
 In fact, we can show that two first-order open formulae $\phi(x_1,..x_n)$ and
 $\psi(x_1,..x_n)$ are intensionally \emph{equivalent}  iff $\diamondsuit \phi(x_1,..x_n)$ and  $\diamondsuit
 \psi(x_1,..x_n)$ are intensionally \emph{equal}. We have that
$I_n(\diamondsuit \phi(x_1,...,x_k))(w') = \|\diamondsuit
\phi(x_1,...,x_k)\|_{\M, w'}\\ =
 \{ (g(x_1),...,g(x_k)) \in \D^k ~| ~g \in \D^{\V}$ and  ${\M}
\models_{w',g}~ \diamondsuit \phi(x_1,...,x_k) \}  \\
 =  \{ (g(x_1),...,g(x_k))  ~| ~g \in \D^{\V}$ and  exists $w$ such that $(w',w)\in \R$\\ and ${\M}
\models_{w,g}~ \phi(x_1,...,x_k) \}  \\
=  \{ (g(x_1),...,g(x_k))  ~| ~g \in \D^{\V}$ and  exists $w$ such
that  ${\M}\models_{w,g}~ \phi(x_1,...,x_k) \}  \\
= \bigcup_{w \in \W} I_n( \phi(x_1,...,x_k))(w) = \bigcup_{w \in \W}
w^*(\phi(x_1,...,x_k))$,\\
that is, the intension of $\diamondsuit \phi(x_1,..x_n)$ is a
\emph{constant} function.\\ Thus,  $\phi(x_1,..x_n)$ and
 $\psi(x_1,..x_n)$ are intensionally \emph{equivalent} if \\$~\bigcup_{w \in \W} I_n(
 \phi(x_1,...,x_k))(w) = \bigcup_{w \in \W} I_n(
 \psi(x_1,...,x_k))(w)$, i.e.,\\ if $I_n(\diamondsuit
 \phi(x_1,...,x_k))(w) = I_n(\diamondsuit
 \phi(x_1,...,x_k))(w)$ for every world $w \in \W$, i.e.,\\ if $\diamondsuit \phi(x_1,..x_n)$ and  $\diamondsuit
 \psi(x_1,..x_n)$ are intensionally \emph{equal}.
 \\$\square$\\
 Another extension of this minimal intensional FOL is of course the
 intensional FOL defined by Bealer in \cite{Beal82}, if we define
 the mapping $is:\W \rightarrow \E$ in Definition \ref{def:intensemant} as the
 Montague-Bealer's
 isomorphism (bijection) between possible worlds and the set of
 extensionalization functions.\\
 Notice that both versions of intensional FOL are modal logics, thus
 we can define two different  logic inferences for them: the \emph{local}
 inference relation $\vdash_w$ and the \emph{global} inference relation
 $\vdash$, as follows:
 \begin{enumerate}
   \item For a given set of logic formulae $\Gamma$ we tell that they
 locally infer the formula $\phi$ in a possible world $w \in \W$, that is,\\ $~\Gamma \vdash_w \phi~~$
 iff
 $~~(\forall$ models $\M)(\forall g)((\forall \psi \in \Gamma).{\M}\models_{w,g}~
 \psi$ implies $ {\M}\models_{w,g}~ \phi)$.
   \item For a given set of logic formulae $\Gamma$ we tell that they
 globally infer the formula $\phi$, that is, $~\Gamma \vdash \phi~~$
 iff
 $~~(\forall$ models $\M)(\forall g)(\forall w \in \W)((\forall \psi \in \Gamma).{\M}\models_{w,g}~
 \psi$ implies $ {\M}\models_{w,g}~ \phi)$.
 \end{enumerate}
 The intensional First-order logic FOL$_{\I}(\Gamma)$ in Definition
 \ref{def:IntensionalFOL}  has \emph{one} unique Kripke model $\M =
 \M_{FOL_{\I}(\Gamma)}$. Thus, we obtain that in this modal intensional logic FOL$_{\I}(\Gamma)$:
\begin{enumerate}
   \item  $~\Gamma \vdash_w \phi~~$ iff $~~\phi$ is true in the Kripke model $\M_{FOL_{\I}(\Gamma)}$ in a given possible world $w  \in \mathfrak{I}_T(\Gamma)$, that
   is, if $~\phi$ is true in the Tarski's model $I_T = w$ of
   $\Gamma$. Thus, this local inference $\vdash_w$ corresponds to the
   derivation of true formulae in a given Tarski model $I_T = w$ of
   $\Gamma$.
   \item  $~\Gamma \vdash \phi~~$  iff
 $~~\phi$ is true in the Kripke model $\M_{FOL_{\I}(\Gamma)}$, that is, iff  $~\Gamma \Vdash
 \phi$. So that the global inference $\vdash$ corresponds to the Tarskian
 logical consequence $\Vdash$ in the standard First-order logic.
 \end{enumerate}
 In the rest of this section we will consider the full homomorphic
 (algebraic) extensions of intensional semantics defined in Definition
 \ref{def:intensemant}. The first step is to define the intensional
 algebra $\A_{int}$ of concepts, analogous to Concept languages as,
 for example, in the case of the Description Logic (DL).\\
 Concept languages steam from semantic networks
\cite{Haye74,Wood75,Haye79} which for a large group of graphical
languages used in the 1970s to represent and reason with conceptual
knowledge. But they did not have a rigorously defined statement as
emphasized by Brachman and Levesque \cite{LeBr85,LeBr87}. After
that, different versions of DL \cite{BCGNP02} with formal semantics
appeared, as a family of knowledge representation formalisms that
represent the knowledge of an application domain by first defining
the relevant \emph{concepts} and \emph{roles}  as a terminology
(TBox) and then the \emph{assertions} (ABox) about named individuals
in terms of this terminology. The concepts denote sets
of individuals, and roles denote binary relationships between individuals.\\
In our approach we will use not only binary, but also general k-ary
relationships between individuals, in order to manage not only unary
(as in DL)  but all k-ary concepts. This approach is similar to
Bealer's intensional algebra, with the difference that our algebra
is not an extension of intensional Boolean algebra as in the
Bealer's work, where the intensional conjunction is extensionally
interpreted by set intersection (here, instead, it is interpreted by
the natural join operations, defined in the FOL extensional algebra
$\A_{\mathfrak{R}}$ in Corollary \ref{coro:intensemant}). Moreover,
we will define only the minimal intensional algebra (with minimal
number of operators), able to support the homomorphic extension of
the intensional mapping $I:\L \rightarrow \D$.
\begin{definition} \textsc{Intensional \textsc{FOL} Algebra:} \label{def:intalgebra} Intensional FOL algebra is a structure \\$~\A_{int} =
~~(\D, Id, Truth, \{conj_{S}\}_{ S \in \P(\mathbb{N}^2)}, neg,
\{exists_{n}\}_{n \in \mathbb{N}})$, $~~$ with
 binary operations  $~~conj_{S}:D_I\times D_I \rightarrow D_I$,
   unary operation  $~~neg:D_I\rightarrow D_I$, and unary
   operations
$~~exists_{n}:D_{I}\rightarrow D_I$, such that for any
extensionalization function $h \in \E$,
and $u \in D_k, v \in D_j$, $k,j \geq 0$,\\
1. $~h(Id) = R_=~$ and $~h(Truth) = \{<>\}$.\\
2. $~h(conj_{S}(u, v)) = h(u) \bowtie_{S}h(v)$, where $\bowtie_{S}$
is the natural join operation defined in Corollary
\ref{coro:intensemant} and $conj_{S}(u, v) \in D_m$ where $m = k + j
- |S|$
 if for every pair $(i_1,i_2) \in S$ it holds that $1\leq i_1 \leq k$, $1 \leq i_2 \leq j$ (otherwise $conj_{S}(u, v) \in D_{k+j}$).\\
3. $~h(neg(u)) = ~\sim(h(u)) = \D^k \backslash (h(u))$,
 where  $~\sim~$ is the operation
defined in Corollary \ref{coro:intensemant} and $neg(u) \in D_k$.\\
 4. $~h(exists_{n}(u)) =
\pi_{-n}(h(u))$, where $\pi_{-n}$ is the operation defined in
Corollary \ref{coro:intensemant} and $exists_n(u) \in D_{k-1}$ if $1
\leq n \leq k$  (otherwise $exists_n$ is the identity function).
\end{definition}
We define the following homomorphic extension of the intensional
interpretation $I:\L \rightarrow \D$:
\begin{enumerate}
  \item The logic formula $\phi(x_i,x_j,x_k,x_l,x_m) \wedge \psi
(x_l,y_i,x_j,y_j)$ will be intensionally interpreted by the concept
$u_1 \in D_7$, obtained by the algebraic expression $~
conj_{S}(u,v)$ where $u = I(\phi(x_i,x_j,x_k,x_l,x_m)) \in D_5, v =
I(\psi (x_l,y_i,x_j,y_j))\in D_4$ are the concepts of the virtual
predicates $\phi, \psi$, relatively, and $S = \{(4,1),(2,3)\}$.
Consequently, we have that for any two formulae $\phi,\psi \in \L$
and a particular  operator $conj_S$ uniquely determined by tuples of
free variables in these two formulae, $I(\phi \wedge \psi) =
conj_{S}(I(\phi),I(\psi))$.
  \item The logic formula $\neg \phi(x_i,x_j,x_k,x_l,x_m)$ will be
intensionally interpreted by the concept $u_1  \in D_5$, obtained by
the algebraic expression $~neg(u)$ where $u =
I(\phi(x_i,\\x_j,x_k,x_l,x_m)) \in D_5$ is the concept of the
virtual predicate $\phi$. Consequently, we have that for any formula
$\phi \in \L$, $~I(\neg \phi) = neg(I(\phi))$.
  \item The logic formula $(\exists x_k) \phi(x_i,x_j,x_k,x_l,x_m)$ will
be intensionally interpreted by the concept $u_1  \in D_4$, obtained
by the algebraic expression $~exists_{3}(u)$ where $u =
I(\phi(x_i,x_j,x_k,x_l,x_m)) \in D_5$ is the concept of the virtual
predicate $\phi$. Consequently, we have that for any formula $\phi
\in \L$ and a particular operator $exists_{n}$ uniquely determined
by the position of the  existentially quantified variable in the
tuple of free variables in $\phi$ (otherwise $n =0$ if this
quantified variable is not a free variable in $\phi$), $~I((\exists
x)\phi) = exists_{n}(I(\phi))$.
\end{enumerate}
\begin{coro} \textsc{Intensional/extensional FOL semantics:} \label{coro:intalgebra}
For any Tarski's interpretation $I_T$ of the FOL, the following
 diagram of homomorphisms commutes:
\begin{diagram}
    &    & \A_{int}~ (concepts/meaning) & &\\
  & \ruTo^{I ~(intensional~int.)} & \frac{Frege/Russell}{semantics}  &\rdTo^{h (extensionaliz.)} &\\
 \A_{FOL}~(syntax)  &    & \rTo_{I_T^*~(Tarski's ~interpretation)} && \A_{\mathfrak{R}} ~(denotation)   \\
\end{diagram}
where $h = is(w)$ where $w = I_T \in \W$ is the explicit possible
world of the minimal intensional first-order logic  in Definition
\ref{def:IntensionalFOL}.
\end{coro}
\textbf{Proof:} The homomorphism of intensional mapping $I$ is
defined by intensional interpretation above. Let us show that also
the isomorphism $is$ between the extensionalization mappings $h$ and
Tarski's interpretations  $I_T$ is uniquely determined in order to
make homomorphic and commutative the diagram above. It can be done
by inductive structural recursion on the length of FOL formulae in
$\L$: for any atom $p_i^k(x_1,...,x_k) \in \L$ we define $is: I_T
\mapsto h$ by requirement that $h(I(p_i^k(x_1,...,x_k))) =
I_T(p_i^k)$. Let us suppose that for any formula $\phi$ with $n$
logic connectives it holds that the mapping $is: I_T \mapsto h$
satisfies requirement that $h(I(\phi)) = I_T^*(\phi)$. Let us show
that it holds also for any logic formula $\phi$ with $n+1$ logic
connectives. It is enough to show it in the case when $\varphi =
\phi
\wedge \psi$ (the other two cases are analogous):\\
$h(I(\varphi)) = h(I(\phi \wedge \psi)) =
h(conj_{S}(I(\phi),I(\psi)))$ (from the homomorphic property of $I$)
$ =_{def} h(I(\phi)) \bowtie_{S} h(I(\psi))$ (from Definition
\ref{def:intalgebra}) $ = I_T^*(\phi) \bowtie_{S} I_T^*(\psi)$ (by
inductive hypothesis) $ = I_T^*(\varphi)$, from the fact that the
same conjunctive formula $\varphi$ is mapped by $I$ into
$conj_{S_1}$ and by $I_T^*$ into $\bowtie_{S_2}$ where $S_1 = S_2$.
\\$\square$\\
This homomorphic diagram formally express the fusion of Frege's and
Russell's semantics \cite{Freg92,Russe05,WhRus10} of meaning and
denotation of the FOL language, and renders mathematically correct
the definition of what we call an "intuitive notion of
intensionality", in terms of which a language is intensional if
denotation is distinguished from sense: that is, if both a
denotation and sense is ascribed to its expressions. This notion is
simply adopted from Frege's contribution (without its infinite
sense-hierarchy, avoided by Russell's approach where there is only
one meaning relation, one fundamental relation between words and
things, here represented by one fixed intensional interpretation
$I$), where the sense contains mode of presentation (here described
algebraically as an algebra of concepts (intensions) $\A_{int}$, and
where sense determines denotation for any given extensionalization
function $h$ (correspondent to a given Traski's interpretaion
$I_T$). More about the relationships between Frege's and Russell's
theories of meaning may be found in the Chapter 7,
"Extensionality and Meaning", in \cite{Beal82}.\\
As noted by Gottlob Frege and Rudolf Carnap (he uses terms
Intension/extension in the place of Frege's terms sense/denotation
\cite{Carn47}), the two logic formulae with the same denotation
(i.e., the same extension for a given Tarski's interpretation $I_T$)
need not have the same sense (intension), thus such co-denotational
expressions are not
\emph{substitutable} in general.\\
In fact there is exactly \emph{one} sense (meaning) of a given logic
formula in $\L$, defined by the uniquely fixed intensional
interpretation $I$, and \emph{a set} of possible denotations
(extensions) each determined by a given Tarski's interpretation of
the FOL as follows from Definition \ref{def:intensemant},
\begin{center}
$~~~ \L ~\longrightarrow_I~ \D ~\Longrightarrow_{h = is(I_T) \& I_T
\in ~\W = \mathfrak{I}_T(\Gamma)}~ \mathfrak{R}$.
\end{center}
Often 'intension' has been used exclusively in connection with
possible worlds semantics, however, here we use (as many others; as
Bealer for example) 'intension' in a more wide sense, that is as an
\emph{algebraic expression} in the intensional algebra of meanings
(concepts) $\A_{int}$ which represents the structural composition of
more complex concepts (meanings) from the given set of atomic
meanings. Consequently, not only the denotation (extension) is
compositional,
but also the meaning (intension) is compositional.\\
Notice that this compositional property holds also for the
generation of subconcepts: for example, given a virtual predicate
$\phi(x_1,..x_n)$ with correspondent concept $I(\phi) \in D_n$, its
subconcept is defined by $I(\phi[x_i/c]) =
I(\phi(x_1,...,x_{i-1},[x_i/c], x_{i+1},...,x_n)) \in D_{n-1}$,
where the i-th free variable of the original virtual predicate is
substituted by a language constant $c$. \\The following
compositional relationship exists between extensions of concepts and
their subconcepts:
\begin{propo} For any extensionalization function $h$  and a virtual predicate $\phi$ with a tuple of free variables $(x_1,...,x_{i-1},x_i,
x_{i+1},...,x_n)$, $n \geq i \geq 1$, it holds
that,\\
$h(I(\phi[x_i/c])) = \\ ~ = ~ \pi_{-i}(\{ (u_1,...,u_{i-1},u_i,
u_{i+1},...,u_n) \in h(I(\phi))~|~u_i = I(c)\})$, $~~$ if $~n \geq 2$;\\
$~ = ~ f_{<>}(\{ (u) \in h(I(\phi))~|~u = I(c)\})$, $~~$ if $~i = n
= 1$.\\
For the sentences we have that for any virtual predicate
$\phi(x_1,...,x_n)$ and an assignment $g$, $~~h(I(\phi/g) = t ~~$
iff $~~(g(x_1),...,g(x_n)) \in h(I(\phi))$.
\end{propo}
\textbf{Proof:} Directly from the homomorphic  diagram of
Frege/Russell's intensional semantics in Corollary
\ref{coro:intalgebra}. Let us consider the
first case when $~n \geq 2$, then:\\
$h(I(\phi[x_i/c])) = I_T^*(\phi[x_i/c])) = \{
(g(x_1),...,g(x_{i-1}), g(x_{i+1}),...,g(x_n)) \in D_{n-1}~|\\~g \in
\D^{\V}$ and $I_T^*(\phi(g(x_1),...,g(x_{i-1}),
I(c),g(x_{i+1}),...,g(x_n))) = t \}  \\ = \pi_{-i}(\{
(g(x_1),...,g(x_{i-1}),g(x_i), g(x_{i+1}),...,g(x_n)) \in
D_{n-1}~|~g \in \D^{\V}$ and $I_T^*(\phi/g) = t$ and $g(x_i) = I(c)
\})\\ = \pi_{-i}(\{ (g(x_1),...,g(x_{i-1}),g(x_i),
g(x_{i+1}),...,g(x_n)) \in I_T^*(\phi)~|~g \in \D^{\V}$ and  $g(x_i)
= I(c) \})\\  = \pi_{-i}(\{ (u_1,...,u_{i-1}, u_i, u_{i+1},...,u_n)
\in I_T^*(\phi)~|~  u_i = I(c) \})\\ = \pi_{-i}(\{ (u_1,...,u_{i-1},
u_i, u_{i+1},...,u_n) \in h(I(\phi))~|~ u_i = I(c) \})$.\\ The other
cases are analogous.
\\$\square$\\
From this proposition it is clear the importance of the homomorphic
extensions of the two-step intensional semantics in Definition
\ref{def:intensemant}. Without this homomorphic commutativity with
the Tarski's interpretations, given by Corollary
\ref{coro:intalgebra}, it will not be able to specify the
interdependence of extensions of correlated concepts in $\D$. Thus,
the homomorphic extension of Frege/Russell's intensional semantics
is not only a meaningful theoretical contribution but also a
necessarily issue in order to be able to define the correct
intensional semantics for the
FOL.\\
The commutative homomorphic diagram in Corollary
\ref{coro:intalgebra} explains in which way the Tarskian semantics
neglects meaning, as if truth in language where autonomous. This
diagram show that such a Tarskian approach, quite useful in logic,
is very approximative. In fact the Tarskian fact "A is a true
sentence" (horizontal arrow in the diagram above with $I_T^*(A) =
t$), is equivalent to "A expresses a true proposition" (where the
proposition is an intensional entity equal to $I(A)$, and its truth
is obtained by extensionalization mapping $h(I(A)) = t$). That is,
the diagram above considers also the theory of truth as a particular
case of the theory of meaning, where we are dealing with
propositions in $D_0 \subset \D$.\\ Because of that, the
intensionality is a strict generalization of the Tarskian theory of
truth that is useful in mathematical logic but inessential to the
semantics for natural language. It explains why the modern
intelligent information retrieval in Web P2P database systems
requires the intensionality, and the application of the general
theory of meaning in the place of the singular Tarskian theory of
truth.

%
\section{Conclusion}
Semantics is the theory concerning the fundamental relations between
words and things. In Tarskian semantics of the FOL one defines what
it takes for a sentence in a language to be true relative to a
model. This puts one in a position to define what it takes for a
sentence in a language to be valid. Tarskian semantics often proves
quite useful in logic. Despite this, Tarskian semantics neglects
meaning, as if truth in language were autonomous. Because of that
the Tarskian theory of truth becomes inessential to the semantics
for more expressive logics, or more 'natural' languages, and it is
the starting point of my investigation about how to provide the
necessary, or minimal, intensionality to the syntax of the FOL.\\
Both, Montague's and Bealer's approaches were useful for this
investigation, but the first is not adequate and explains why we
adopted two-step intensional semantics (intensional interpretation
with the set of extensionalization functions), and the second
consider that the intensionality is exclusive consequence of
"intensional abstraction". First, we show that not all modal
predicate logics are intensional logics but only a strict subset of
them are intensional. Also the set of pure extensional predicate
logics is the strict subset of modal predicate logics.\\
We defined a  \emph{modal} FOL$_{\K}$ logic where the quantifiers
are interpreted as modal operators, and we have shown that such a
modal predicate logic (heaving the same syntax as ordinary FOL) with
Kripke's possible world semantics is pure \emph{extensional} logic
as is FOL with standard Tarskian semantics. We show that the
transformation of this predicate modal logic FOL$_{\K}$ into FOL,
 by using correspondence modal theory,
is impossible, from the fact that by transformation of the modal formulae we obtain the second-order formulae (because the possible worlds are the functions
of assignments). In the same way, the transformation of the \emph{intensional} first-order logic FOL$_{\I}$ into FOL is impossible (the set of possible worlds
are the functions of Tarski's models of the standard FOL with a set of assumptions $\Gamma$).\\
We have shown that minimal intensional enrichment of the FOL (which
does not change the syntax of the FOL) is obtained by adopting the
PRP theory, that is a theory of properties, relations, and
propositions for the domain $\D$ of the FOL, and by adopting the
two-step intensional interpretation. The set of possible worlds of
this 'minimal' intensional logic FOL$_{\I}$ is the set of Tarski's
models of the standard FOL with a set of assumptions $\Gamma$, with
the intensionality equal to Montague's point of view of the meaning.
The global logical inference relation of this intensional
first-order logic FOL$_{\I}$ is equal to the standard Tarskian
logical consequence relation of
the FOL.\\
At the end of this work we defined an intensional algebra and an
extensional algebra (different from standard cylindric algebras for
the FOL), and the commutative homomorphic diagram between them, in
Corollary \ref{coro:intalgebra}, that express the generalization of
the Tarskian theory of truth for the FOL into the Frege/Russell's
theory of meaning in this minimal intensional enrichment of the FOL.


\bibliographystyle{IEEEbib}
\bibliography{medium-string,krdb,mydb}



%
\end{document}